\documentclass{bmcart}

\usepackage{amsthm,amsmath}
\RequirePackage{natbib}
\usepackage[utf8]{inputenc} 
\usepackage{amssymb}
\usepackage{graphicx}
\usepackage{bm}

\startlocaldefs
\let\jnl@style=\rmfamily
\def\ref@jnl#1{{\jnl@style#1}}%
\newcommand\aj{\ref@jnl{AJ}}
\newcommand\apj{\ref@jnl{ApJ}}%
\newcommand\apjl{\ref@jnl{ApJ}}%
\newcommand\mnras{\ref@jnl{MNRAS}}%
\newcommand\na{\ref@jnl{New A}}%
\newcommand\nat{\ref@jnl{Nature}}%
\newcommand\pasj{\ref@jnl{PASJ}}%
\newcommand\aap{\ref@jnl{A\&A}}%

\endlocaldefs

\begin{document}

\begin{frontmatter}

\begin{fmbox}
\dochead{Research}


\title{GPU-Enabled Particle-Particle Particle-Tree Scheme for Simulating Dense Stellar Cluster System}


\author[
   addressref={aff1,aff2},                   
   email={masaki.iwasawa@riken.jp}   
]{\inits{MI}\fnm{Masaki} \snm{Iwasawa}}
\author[
   addressref={aff2},
   email={spz@strw.leidenuniv.nl}
]{\inits{SPZ}\fnm{Simon} \snm{Portegies Zwart}}
\author[
   addressref={aff1,aff3},
   email={makino@mail.jmlab.jp}
]{\inits{JM}\fnm{Junichiro} \snm{Makino}}


\address[id=aff1]{
  \orgname{RIKEN Advanced Institute for Computational Science}, 
  \street{Minatojima-minamimachi, Chuo-ku},                     %
  \city{Kobe},                              
  \cny{Japan}                                    
}

\address[id=aff2]{%
  \orgname{Sterrewacht Leiden},
  \street{P.O. Box 9513, 2300 RA},
  \city{Leiden},
  \cny{The Netherlands}
}

\address[id=aff3]{%
  \orgname{Earth-Life Science Institute, Tokyo Institute of Technology},
  \street{Ookayama, Meguro-ku},
  \city{Tokyo},
  \cny{Japan}
}



\end{fmbox}


\begin{abstractbox}

\begin{abstract} 

We describe the implementation and performance of the ${\rm P^3T}$
(Particle-Particle Particle-Tree) scheme for simulating dense stellar
systems. In ${\rm P^3T}$, the force experienced by a particle is split
into short-range and long-range contributions.  Short-range forces are
evaluated by direct summation and integrated with the fourth order
Hermite predictor-corrector method with the block timesteps. For
long-range forces, we use a combination of the Barnes-Hut tree code
and the leapfrog integrator. The tree part of our simulation
environment is accelerated using graphical processing units (GPU),
whereas the direct summation is carried out on the host CPU. Our code
gives excellent performance and accuracy for star cluster simulations
with a large number of particles even when the core size of the star
cluster is small.

\end{abstract}




\end{abstractbox}
%

\end{frontmatter}




\section{Background}
\label{Sect:Background}

Direct $N$-body simulation has been the most useful tool for the study
of the evolution of collisional stellar systems such as star clusters
and the center of the galaxy \cite{1963MNRAS.126..223A}. The force
calculations, of which the cost is $O(N^2)$, are the most
compute-intensive part of direct $N$-body simulations. Barnes and Hut
\cite{1986Natur.324..446B} developed a scheme which reduces the
calculation cost to $O(N{\rm log}N)$ by constructing the tree
structure and evaluating the multipole expansions. Dehnen
\cite{2002JCoPh.179...27D, 2014ComAC...1....1D} developed a scheme to
reduce the calculation cost to $O(N)$ by combining the fast multipole
method \cite{1987JCoPh..73..325G} and the tree code.  Recently, the
graphical processing units (GPU), which is a device originally
developed for rendering the graphical image, began to be used for
scientific simulations. The tree code is also implemented on GPUs and
it is much faster than that on CPUs \cite{2010ProCS...1.1119G,
  2012JCoPh.231.2825B}. B\'edorf et
al. \cite{Bedorf:2014:PGT:2683593.2683600} parallelized the tree code
on GPUs and showed good scalability up to 18600 GPUs. They also
simulated the Milky Way Galaxy with $N$ of up to 242 billion and
reported that the average calculation time per iteration on 18600 GPUs
was 4.8 seconds.

The tree schemes are widely used for collisionless system simulations.
However, for collisional system simulations, the use of the tree code
has been very limited. One reason might be that a collisional stellar
system spans a wide range in timescales. Thus it is essential that
each particle has its own integration timestep. This scheme is called
the individual timestep or the block timestep
\cite{1986LNP...267..156M}. However, when we use the tree code and the
block timestep together, the tree structure is reconstructed at every
block timestep, because the positions of integrated particle are
updated. The cost of the usual complete reconstruction of the tree is
$O(N{\rm log}N)$ and not negligible.

To reduce the cost of the reconstruction of the tree, McMillan and
Aarseth \cite{1993ApJ...414..200M} introduced local reconstruction of
tree. They demonstrated a good performance, but there seems to be no
obvious way to parallelize their scheme.

Recently, Oshino et al. \cite{2011PASJ...63..881O} introduced another
approach to combine the tree code and the block timesteps which they
called the ${\rm P^3T}$ scheme. This scheme is based on the idea of
Hamiltonian splitting \cite{1991CeMDA..50...59K, 1991AJ....102.1528W,
  1998AJ....116.2067D, 1999MNRAS.304..793C, 2003MNRAS.346..924B,
  2007PASJ...59.1095F, 2011NewA...16..445M}. In the ${\rm P^3T}$
scheme, the Hamiltonian of the system is split into short-range and
long-range parts and they are integrated with different
integrators. The long-range part is evaluated with the tree code and
is integrated using the leapfrog scheme with a shared timestep. The
short range part is evaluated with direct summation and integrated
using the fourth-order Hermite scheme \cite{1992PASJ...44..141M} with
the block timesteps. They investigated the accuracy and the
performance of the ${\rm P^3T}$ scheme for planetary formation
simulations and showed that the ${\rm P^3T}$ scheme achieves high
performance.

In this paper, we present the implementation of the ${\rm P^3T}$
scheme on GPUs and report its accuracy and performance for star
cluster simulations. We found that the ${\rm P^3T}$ scheme
demonstrates a very good performance for star cluster simulations,
even when the core of the cluster becomes small.

The structure of this paper is as follows. In section 2, we briefly
describe the ${\rm P^3T}$ scheme. In section 3, we report the accuracy
and performance of the ${\rm P^3T}$ scheme. We summarize these results
in section 4.

\section{Methods} 
\label{Sect:Method} 
\subsection{Formulation} 

In this section, we describe the ${\rm P^3T}$ scheme. The Hamiltonian
$H$ of a gravitational $N$-body system is given by
\begin{eqnarray} 
H &=& \sum^N_i \frac{|{\bm p}_i|^2}{2m_i} - \sum^N_{i}
\sum^N_{i<j}\frac{Gm_i m_j}{s_{ij}}, \\ s_{ij} &=& \sqrt{|{\bm
q}_{ij}|^2 + \epsilon^2}, \\ {\bm q}_{ij} &=& {\bm q}_i - {\bm q}_j,
\end{eqnarray} 
where ${\bm p}_i$, $m_i$ and ${\bm q}_i$ are momentum,
mass and position of the particle $i$, respectively. To avoid the
singularity of the $1/r$ potential, we use the Plummer softening
$\epsilon$ \cite{1963MNRAS.126..223A}. With the ${\rm P^3T}$ scheme,
$H$ is split into $H_{{\rm hard}}$ and $H_{{\rm soft}}$ as follows
\cite{2011PASJ...63..881O}: 

\begin{eqnarray} H &=& H_{\rm hard} +
H_{\rm soft}, \\ H_{\rm hard} &=& \sum_i^N \frac{|{\bm p}_i|^2}{2m_i}
- \sum^N_{i} \sum_{i<j}^N\frac{m_im_j}{s_{ij}}\left[
1-W(s_{ij})\right], \\ H_{\rm soft} &=& -\sum^N_{i}
\sum_{i<j}^N\frac{m_i m_j}{s_{ij}}W(s_{ij}).  
\end{eqnarray} 

Here $W(s_{ij})$ is a smooth transition function. A suitable form of
$W(s_{ij})$ should be zero when a distance between two particles is
smaller than the inner cutoff radius $r_{\rm in}$ and should be unity
if the distance is larger than the outer cutoff radius $r_{\rm
cut}$. This splitting is introduced by Chambers
\cite{1999MNRAS.304..793C} to avoid undesirable energy error from
close encounters between particles. Similar splitting has been used
with ${\rm P^3M}$ (Particle-Particle Particle-Mesh) scheme, in which
the long-range part of the interaction is evaluated by using FFT
\cite{1981csup.book.....H}.

Forces derived from $H_{\rm hard}$ and $H_{\rm soft}$ are given by
\begin{eqnarray}
{\bm F}_{\rm hard,i} &=& - \frac{\partial H_{\rm hard}}{\partial {\bm q}_i} =  - \sum_{j \neq i}^N\frac{m_im_j}{s_{ij}^3}(1-K(s_{ij})){\bm q}_{ij}, \\
{\bm F}_{\rm soft,i} &=& - \frac{\partial H_{\rm soft}}{\partial {\bm q}_i} = - \sum_{j \neq i}^N\frac{m_im_j}{s_{ij}^3}K(s_{ij}){\bm q}_{ij}, \\
K(s_{ij}) &=& W(s_{ij}) - s_{ij}\frac{d W(s_{ij})}{d s_{ij}}.
\end{eqnarray}
We call $K(s_{ij})$ the cutoff function. 

The tree algorithm is used for the evaluation of ${\bm F}_{\rm soft,i}$
to reduce the calculation cost.

The formal solution of the equation of motion for the phase space
coordinate ${\bm w} = ({\bm q}, {\bm p}) $ at time $t+\delta t$ for
the given Hamiltonian $H$ is
\begin{equation}
{\bm w}(t+\delta t) = e^{\delta t \{, H\}} {\bm w}(t) = e^{\delta t \{, H_{\rm soft} + H_{\rm hard}\}} {\bm w}(t).
\end{equation}
Here the braces $\{,\}$ stand for the Poisson bracket. In the ${\rm P^3T}$
scheme, we use the second order approximation;
\begin{equation}
{\bm w}(t+\delta t) = e^{\delta t /2 \{, H_{\rm soft}\}}e^{\delta t \{, H_{\rm hard}\}}e^{\delta t /2 \{, H_{\rm soft}\}} {\bm w}(t) + {\Large O}(\delta t ^3).
\end{equation}
Here, the formal solution for the $H_{\rm soft}$ term is the simple
velocity kick, since $H_{\rm soft}$ contains the potential only.  We
numerically integrate the $H_{\rm hard}$ term, since it cannot be
solved analytically. We use the fourth-order Hermite scheme with the
block timestep \cite{1992PASJ...44..141M}. The fourth-order integrator
requires $K(s_{ij})$ to be three-times differentiable with respect to
position. We use the following formula:

\begin{eqnarray}
K(x) &=& \left\{
\begin{array}{ll}
0 &(x < 0) \\
-20x^7 + 70x^6 - 84x^5 +35x^4 &( 0 \leq x < 1) \\
1 &( 1 \leq x) \\
\end{array} \right. ,\\
x &=& \frac{y-\gamma}{1-\gamma}, \\
y &=& \frac{s_{ij}}{r_{\rm cut}}, \\
\gamma &=& \frac{r_{\rm in}}{r_{\rm cut}}.
\end{eqnarray}

This $K(x)$ is the lowest-order polynomial which satisfies the requirement
that derivatives up to the third order is zero for $x=0$ and $1$
(i.e. The highest-order term of the lowest-order polynomial is the
seventh, because there are eight boundary conditions at $x=0$ and
$x=1$).

In figure \ref{Fig:Ky}, we plot $K(y)$ (top panel) and forces (bottom
panel) with $\gamma=0.1$. According to \cite{2011PASJ...63..881O,
  1999MNRAS.304..793C}, $K(y)$ with $\gamma=0.1$, is smooth enough to
be integrated. Thus, for all calculations, we use $\gamma=0.1$. The
functional form of $W(y;\gamma)$ is given by

\begin{eqnarray}
W(y; \gamma) &=& \left\{
\begin{array}{ll}
\frac{7(\gamma^6 - 9\gamma^5 + 45\gamma^4 - 60\gamma^3{\rm log}\gamma - 45\gamma^2 + 9\gamma - 1)}{3(\gamma-1)^7}y & (y < \gamma) \\
G(y; \gamma) + (1-G(1;\gamma))y & ( \gamma \leq y < 1) \\
1 &  ( 1 \leq y) \\
\end{array} \right. ,\\
G(y;\gamma) &=& \left( -10/3y^7 + 14(\gamma+1)y^6 - 21(\gamma^2+3\gamma+1)y^5 \right. \nonumber \\
&& + ( 35(\gamma^3+9\gamma^2+9\gamma+1)/3 )y^4 - 70(\gamma^3+3\gamma^2+\gamma)y^3 \nonumber \\
&& + 210(\gamma^3+\gamma^2)y^2 - 140\gamma^3y{\rm log}(y) \nonumber \\
&& \left. + (\gamma^7-7\gamma^6+21\gamma^5-35\gamma^4) \right) / (1-\gamma)^7.
\end{eqnarray}

With the ${\rm P^3T}$ scheme, the time integration proceeds as follows
\begin{enumerate}

\item At time $t$, by using the tree code, calculate the acceleration
  due to $H_{\rm soft}$, ${\bm a}_{{\rm soft},i}$, and construct a list
  of all particles which come within $r_{\rm cut}$ from particle $i$
  for $\Delta t_{\rm soft}$. Here, $\Delta t_{\rm soft}$ is the
  timestep for the soft Hamiltonian.

\item Update the velocities of all particles with ${\bm v}_{{\rm
  new},i}={\bm v}_{{\rm old},i}+(1/2)\Delta t_{\rm soft} {\bm a}_{{\rm
  soft},i}$.

\item Integrate all particles to time $t+\Delta t_{\rm soft}$ under
  $H_{\rm hard}$ , using the neighbour list and the fourth order Hermite
  integrator with the block timesteps.

\item Calculate the acceleration due to $H_{\rm soft}$ at new time
  $t+\Delta t_{\rm soft}$ and update the velocity

\item Go back to step 2.

\end{enumerate}

For the timestep criterion for the block timestep, we use the
following form \cite{2011PASJ...63..881O}.

\begin{eqnarray}
\Delta t_{i} &=& \min \left(  \eta \sqrt{ \frac{\sqrt{|{\bm a}_{i}^{(0)}|^2+a_{0}^2}|{\bm a}_{i}^{(2)}| 
    + |{\bm a}_{i}^{(1)}|^2}{|{\bm a}_{i}^{(0)}||{\bm a}_{i}^{(3)}| + |{\bm a}_{i}^{(2)}|^2}  }, \, \Delta t_{\rm max} \right), \label{eq:timestep:0} \\
a_{0} &=& \alpha \frac{m}{r_{\rm cut}^2}.
\label{eq:timestep:1}
\end{eqnarray}

Here $\eta$ is the accuracy parameter of the timestep and its typical
value is 0.1. $\Delta t_{\rm max}$ is the maximum timestep which
should be smaller than $\Delta t_{\rm soft}$, ${\bm a}_{i}^{(n)}$ is
the $n$th time derivative of the acceleration of particle $i$, $a_0$
is a constant introduced to prevent $\Delta t_{i}$ from becoming too
small when the distance to the nearest neighbor is close to $r_{\rm
  cut}$ and $\alpha$ is a parameter to control $a_0$. In this case,
the acceleration from $H_{\rm hard}$ becomes very small and there is
no need to use very small $\Delta t_{i}$. According to
\cite{2011PASJ...63..881O}, when we chose $\alpha \le 1$, $\alpha$
hardly affects the energy error. Thus we set $\alpha = 0.1$ for all
simulations.

In our Hermite implementation, ${\bm a}_{i}^{(2)}$ and ${\bm
  a}_{i}^{(3)}$ are derived using interpolation of ${\bm a}_{i}^{(0)}$
and ${\bm a}_{i}^{(1)}$, and as a consequence we cannot use equation
(\ref{eq:timestep:0}) for the first step. We use:
\begin{equation}
\Delta t_{i} = \min \left( \eta_{s} \sqrt{ \frac{|{\bm
      a}_{i}^{(0)}|^2+a_{0}^2}{|{\bm a}_{i}^{(1)}|^2} }, \, \Delta t_{\rm
  max} \right).
\label{eq:timestep:2}
\end{equation}
This criterion dose not contain the 2nd and 3rd time derivatives of
the acceleration. To prevent the timestep derived by equation
(\ref{eq:timestep:2}) from becoming too large, we set $\eta_{s}$ to be
the one-tenth of $\eta$ for all simulation in this paper.

We summarize all accuracy parameters in table 1.

\subsection{Implementation on GPUs}\label{Sect:GPU}

Even with the Barnes-Hut tree algorithm, obtaining ${\bm F}_{\rm
  soft,i}$ is still costly and dominates the total calculation time
\cite{2011PASJ...63..881O}. To accelerate this part, we use GPUs, by
modifying the {\tt sequoia} library (B\'edorf, Gaburov and Portegies
Zwart, submitted to ComAC), on which the high-performance tree code
for parallel GPUs {\tt Bonsai} \cite{2012JCoPh.231.2825B} is based.
Our library calculates the long range forces on all particles, ${\bm
  F}_{\rm soft,i}$ by the Barnes-Hut tree algorithm (up to the
quadrupole moment). On the other hand, we calculate ${\bm F}_{\rm
  hard,i}$ on the host computer. The library also returns, for each
particle, the list of particles within the distance $h$ from it. We
use this list of neighbors to calculate ${\bm F}_{\rm hard,i}$. The
value of $h$ should be sufficiently larger than $r_{\rm cut}$ to
guarantee that the particles which are not on the list of the
neighbors of particle $i$ do not enter the sphere of the radius
$r_{\rm cut}$ around particle $i$ during the time interval $\Delta
t_{\rm soft}$.

We call the sphere with a radius of $r_{\rm cut}$ the neighbor sphere
and the shell between the sphere with a radius of $h$ and the neighbor
sphere the buffer shell. The particles of which the nearest neighbor
is outside the sphere with radius $h$ are considered isolated and the
particles on the list of neighbors are considered neighbor
particles. We denote the width of the buffer shell as $\Delta r_{\rm
buff}$ (i.e. $h=r_{\rm cut}+\Delta r_{\rm buff}$).

The compute procedures of our implementation of the ${\rm P^3T}$
scheme on GPU is as follows:
\begin{enumerate}
\item Evaluate long range forces on all particles ${\bm F}_{\rm soft,i}$
  using GPU.
\item Particles are divided into two groups; isolated and
  non-isolated, by using the neighbour list made on GPU.
\item For non-isolated particles, ${\bm F}_{\rm hard,i}$ are  calculated on the host computer.
\item All particles receive a velocity kick through ${\bm F}_{\rm
  soft,i}$ for $\Delta t_{\rm soft}/2$.
\item Isolated particles are drifted by ${\bm r}_{i} \leftarrow {\bm
  r}_{i}+\Delta t_{\rm soft}{\bm v}_{i}$.
\item Non-isolated particles are integrated with the fourth-order
  Hermite scheme for $\Delta t_{\rm soft}$.
\item Evaluate ${\bm F}_{\rm soft,i}$ and make the neighbour list in the same way as in step 1-2.
\item All particles obtain the velocity kick again for $\Delta t_{\rm
  soft}/2$.
\item go back to step 3.
\end{enumerate}

\section{Results}

\subsection{Accuracy and Performance}
\label{Sect:accuracy_and_performance}

We performed a number of test calculations using the ${\rm P^3T}$
scheme on GPUs, to study its accuracy and performance. In this
section, we describe the result of these tests. For most of them we
adopted a Plummer model \cite{1911MNRAS..71..460P} with 128K
(hereafter K=$2^{10}$) equal-mass particles as the initial
condition. We use the so-called $N$-body unit or Heggie unit, in which
total mass M=1, the gravitational constant G=1 and total energy
$E=-1/4$ \cite{1986LNP...267..233H}. To avoid the singularity of the
gravitational potential, we use the Plummer softening and set
$\epsilon = 4/N$. Since this value is a typical separation of a hard
binary in the $N$-body unit, we can follow the evolution of the system
up to the moment of the core collapse.

Note, in this paper, we use the energy errors as an indicator of the
accuracy of the scheme. However, energy conservation dose not
guarantee accuracy of simulations (though, it is necessary). Thus we
will perform realistic simulations in section 3.2 and check the
statistical character of stellar systems by comparing the results with
the Hermite scheme, which is widely used in collisional stellar system
simulations. As we will see later, for simulations of the core
collapse of the star cluster, when the relative energy error is
$\lesssim 10^{-3}$ at the moment of the core collapse, the behavior of
the core collapse with the ${\rm P^3T}$ scheme agreed with that with
the Hermite scheme very well.

\subsubsection{Accuracy}
\label{Sect:accuracy}

With the ${\rm P^3T}$ scheme, we have six accuracy parameters. In
sections \ref{Sect:accuracy:1}-\ref{Sect:accuracy:3}, we discuss how
each parameter controls the accuracy of the ${\rm P^3T}$ scheme. In
section \ref{Sect:accuracy:long}, we describe the accumulation of the
energy error in a long-term integration. To measure energy errors
accurately, we calculate potential energies by the direct summation
instead of the tree code for all runs in this paper.

\paragraph{Effect of $r_{\rm cut}$, $\Delta t_{\rm soft}$ and $\theta$}
\label{Sect:accuracy:1}

In figure \ref{Fig:r-de_t-de}, we present the maximum relative energy
error $|\Delta E_{\rm max}/E_0|$ over 10 $N$-body time units as a
function of $r_{\rm cut}$ and $\Delta t_{\rm soft}$ for several
different values of the opening criterion of the tree, $\theta$.  Here
$\Delta E_{\rm max}$ is the maximum energy error and $E_0$ is the
initial energy. We chose $\eta=0.1$, $\Delta t_{\rm max}=\Delta t_{\rm
  soft}/4$ and $\Delta r_{\rm buff}=3\sigma\Delta t_{\rm soft}$, where
$\sigma$ is the global three dimensional velocity dispersion and we
adopt $\sigma=1/{\sqrt 2}$.

We can see that the error is smaller for smaller $\theta$, smaller
$\Delta t_{\rm soft}$, or larger $r_{\rm cut}$. Roughly speaking, the
error depends on two terms, $\Delta t_{\rm soft}/r_{\rm cut} \sigma$
and $\theta$. If $\Delta t_{\rm soft}/r_{\rm cut} \sigma$ is large, it
determines the error. In this regime, the error is dominated by the
truncation error of the leapfrog integrator. If it is small enough,
$\theta$ determines the error, in other words, the tree force error
dominates the total error. Even for a very small value of $\theta$
like 0.2, the tree force error dominates if 
$\Delta t_{\rm soft}/r_{\rm cut} \sigma \lesssim 0.05$.

In figure \ref{Fig:th-de}, we plot the maximum energy error as a
function of $\theta$. We use the same $\eta$, $\Delta t_{\rm max}$ and
$\Delta r_{\rm buff}$ as in figure \ref{Fig:r-de_t-de}. For the runs
with $r_{\rm cut}=1/256$ and $\Delta t_{\rm soft} =1/512$, the energy
error does not drop below $10^{-6}$ because the error of the leapfrog
integrator is larger than the tree force error. In an chaotic system
like the model used in our simulations such energy error is sufficient
to warrant a scientifically reliable result
\cite{2014ApJ...785L...3P}. On the other hand, for the run with
$r_{\rm cut}=1/128$ and $\Delta t_{\rm soft} =1/1024$, integration
error is smaller than the tree force error.

\paragraph{Effect of $\Delta r_{\rm buff}$}
\label{Sect:accuracy:2}

In figure \ref{Fig:dr-de}, we show the maximum relative energy error
as a function of $\Delta r_{\rm buff}$ for the runs with $\Delta
t_{\rm max}=\Delta t_{\rm soft}/4$, $\eta=0.1$, $\theta=0.2$, for
($\Delta t_{\rm soft}$, $r_{\rm cut}$) = (1/512, 1/128) and (1/1024,
1/256). The energy error is almost constant for $\Delta r_{\rm buff}
\gtrsim 2\Delta t_{\rm soft} \sigma$, which indicates that the energy
error for $\Delta r_{\rm buff} < 2\Delta t_{\rm soft} \sigma$ is
caused by particles that are initially outside the buffer shell (with
radius $r_{\rm cut}+\Delta r_{\rm buff}$) and plunge into the
neighbour sphere (with radius $r_{\rm cut}$) during the timestep
$\Delta t_{\rm soft}$. We can prevent this by adopting $\Delta r_{\rm
  buff} \gtrsim 2\Delta t_{\rm soft} \sigma$.

\paragraph{Effect of $\Delta t_{\rm max}$ and $\eta$}
\label{Sect:accuracy:3}

The maximum relative energy errors over 10 $N$-body time units are
shown in the top panel of figure \ref{Fig:eta-de_eta-nstep} as a
function of $\eta$ and the number of steps for the Hermite part (per
particle per unit time, $N_{\rm step}$) are presented in the bottom
panel. The energy errors go down as $\eta$ decrease until $\eta \sim
0.2$. For $\eta \lesssim 0.2$, the errors hardly depend on $\Delta
t_{\rm max}$.

\paragraph{Long term integration}
\label{Sect:accuracy:long}

In figure \ref{Fig:Eerr-long}, we show the time evolution of the
relative energy error until $T$=500. We compare the accuracy of our
${\rm P^3T}$ scheme with two other schemes, the direct fourth-order
Hermite scheme and the leapfrog scheme with the Barnes-Hut tree
code. The calculations with the direct Hermite scheme are performed by
using the {\tt Sapporo} library on GPU \cite{2009NewA...14..630G}, and
the calculations with the leapfrog scheme are performed by using the
{\tt Bonsai} library on GPU \cite{2012JCoPh.231.2825B}. The energy
error of the ${\rm P^3T}$ scheme behaves like a random walk whereas
that of the leapfrog and the Hermite schemes grow monotonically. In
the right-hand panels of figure \ref{Fig:Eerr-long}, we show the same
evolution of the error as in the left panels, but time is plotted with
a logarithmic scale. This allows us to realize that the error growth
of Hermite and tree schemes are linear, whereas the error in the ${\rm
  P^3T}$ scheme grows as $\propto T^{1/2}$.  This latter
proportionality is caused by the short-term error of the ${\rm P^3T}$
scheme, which is dominated by the randomly changing tree-force
error. For long-term integration the ${\rm P^3T}$ scheme conserves
energy better than the Hermite or leapfrog schemes.

\subsubsection{Calculation cost}
\label{Sect:Method:cost}
In this section, we discuss the calculation cost of the ${\rm P^3T}$
scheme and its dependence on the number of particles $N$, required
accuracy, and other parameters.

We first construct a simple theoretical model of the dependence of the
calculation cost on parameters of the integration scheme such as $N$,
$\Delta t_{\rm soft}$, $\theta$ and $r_{\rm cut}$ in section
\ref{Sect:Method:cost:model}. In section \ref{Sect:Method:cost:set} we
derive the optimal set of parameters from the model and compare this
model with the result of the numerical tests. We found that the
calculation cost per unit time is proportional to $N^{4/3}$.

\paragraph{Theoretical model}\label{Sect:Method:cost:model}
The calculation cost for the force evaluations in ${\rm P^3T}$ is
split into the tree part and the Hermite part. For the tree part, the
calculation cost of evaluating forces for all particles per tree step
is proportional to $\Large{O}(\theta^{-3}N{\rm log}N)$. Since we use
constant timestep for the tree part, the calculation costs of the
integration of particles per unit time for the tree part is
proportional to $\Large{O}\left( \theta^{-3}N{\rm log}N/\Delta t_{\rm
  soft} \right)$.

For the Hermite part, since each particle has its own neighbour
particles and timesteps, the number of interactions for all particles
per unit timstep is given by
\begin{eqnarray}
N_{\rm int, hard} &  = & \sum_i^N N_{{\rm ngh},i} N_{{\rm step},i}\\ 
               &\sim& \sum_i^N 4\pi/3 (r_{\rm cut}+\Delta r_{\rm
                      buff})^3n_i \langle \Delta t_i \rangle ^{-1} \\
               &\propto& N^2(r_{\rm cut}+\Delta r_{\rm buff})^3 \langle
                         \langle \Delta t\rangle \rangle ^{-1},
\end{eqnarray}
Here $N_{{\rm ngh},i}$ is the number of the neighbour particles around
particle $i$, $N_{{\rm step},i}$ is the number of timesteps required
to integrate particle $i$ for one unit time, $n_i$ is the local
density around particle $i$, $\langle \Delta t_i\rangle$ is the
average timestep of particle $i$ over one unit time and $\langle
\langle \Delta t\rangle\rangle$ is the average of $\langle \Delta
t_i\rangle$ over all particles. Here we assume $n_i$ is constant
within the radius of $r_{\rm cut}+\Delta r_{\rm buff}$ around particle
$i$.

Next we express the $\langle \langle \Delta t \rangle \rangle$ as a
function of $N$ and $r_{\rm cut}$. To simplify the discussion, we
define the timestep of the particle through the relative position and
velocity from its nearest neighbour particle; $\langle \langle \Delta
t \rangle \rangle \propto r_{\rm NN}/v_{\rm NN}$, where $r_{\rm NN}$
and $v_{\rm NN}$ are the relative position and the velocity of the
nearest neighbour particle. We can replace $v_{\rm NN}$ to the
velocity dispersion $\sigma$. Thus average timestep is given by
\begin{equation}
\langle\langle \Delta t \rangle\rangle \propto r_{\rm NN}/v_{\rm NN} \sim r_{\rm NN}/\sigma.
\label{eq:timestep} 
\end{equation}

To further simplify the derivation we assume that the number density
of particles in the system is uniform. If $r_{\rm cut}$ is larger than
the mean inter-particle distance $\langle r \rangle$ (i.e. if most
particles have neighbour particles), the average timestep is roughly
given by
\begin{equation}
\langle \langle \Delta t \rangle \rangle \sim \min \left( \eta
\frac{R}{\sigma}N^{-1/3}, \, {\Delta t_{\rm max}} \right),
\label{eq:dtave1}
\end{equation}
where $R$ is the typical size of the system. In this case, the average
timestep depend only on $N$ (dose not depend on $r_{\rm cut}$).

If $r_{\rm cut}$ is small compared to $\langle r \rangle$, most
particles are isolated and most of the non-isolated particles have
only one neighbour particle. In this case, $\langle\langle \Delta t
\rangle \rangle$ is given by
\begin{equation}
\langle\langle \Delta t \rangle \rangle \sim  \min \left( \eta \frac{r_{\rm cut}}{\sigma}, \, {\Delta t_{\rm max}} \right).
\label{eq:dtave2}
\end{equation}

In figure \ref{Fig:rcut-nstep} we show the number of steps per
particle per unit time $N_{\rm step}$ for a plummer sphere as a
function of $N$ (top panel) and as a function of $r_{\rm cut}$ (bottom
panel). In the top panel, we can see that $N_{\rm step}$ is roughly
proportional to $N^{1/3}$ for large $N$ (i.e. $\langle r \rangle$ is
small). On the other hand when $N$ is small $N_{\rm step}$ is almost
constant because $\langle r \rangle$ is large [see equation
  (\ref{eq:dtave2})].

The bottom panel of figure \ref{Fig:rcut-nstep} shows that all curves
eventually approach to constant values for both of large and small
$r_{\rm cut}$. For large $r_{\rm cut}$, the timesteps of the
non-isolated particles are determined by $N$, not by $r_{\rm cut}$
[see equation (\ref{eq:dtave1})], whereas for small values of $r_{\rm
  cut}$ the non-isolated particles have a timesteps $\Delta t_{\rm
  max}$. This is because most neighbouring particles are in the buffer
shell and not in the neighbour sphere.  For runs with $\Delta t_{\rm
  soft}$=1/2048, 1/1024 and 1/512, we can see bumps of $N_{\rm step}$ at
$r_{\rm cut} \sim 1/512$ due to the dependence on $r_{\rm
  cut}$ shown in equation (\ref{eq:dtave2}).

Using above discussions, the number of interactions for all particles
per unit time of the Hermite part $N_{\rm int, hard}$ and the tree
part $N_{\rm int, soft}$ are given by
\begin{eqnarray}
N_{\rm int, hard} &\propto& \left\{
\begin{array}{ll}
N^{7/3}(r_{\rm cut}+\Delta r_{\rm buff})^3 \quad  & ({\rm for} \quad r_{\rm cut} \gg \langle r \rangle) \\
N^2(r_{\rm cut}+\Delta r_{\rm buff})^3 \quad  & ({\rm for} \quad r_{\rm cut} \ll \langle r \rangle)\\
\end{array} \right. , \\
N_{\rm int, soft} &\propto& \theta^{-3}N {\rm log}N / \Delta t_{\rm soft},
\end{eqnarray}

\paragraph{Optimal set of accuracy parameters}\label{Sect:Method:cost:set}

In this section, we derive the optimal values of $r_{\rm cut}$ and
$\Delta t_{\rm soft}$ from the point of view of the balance of the
calculation costs between the tree and the Hermite parts, in other
words we express $r_{\rm cut}$ and $\Delta t_{\rm soft}$ as functions
of $N$ such that $N_{\rm int, hard}/N_{\rm int, soft}$ is independent
of $N$. Following the discussion in section \ref{Sect:accuracy:1} and
because the energy errors can be controlled through $\Delta t_{\rm
  soft}/r_{\rm cut}$, $r_{\rm cut}$ should be proportional to $\Delta
t_{\rm soft}$. From section \ref{Sect:accuracy:2}, $\Delta r_{\rm
  buff}$ should be also proportional $\Delta t_{\rm soft}$.

The requirements are met for $N_{\rm int, hard} \propto N^{7/3}(r_{\rm
  cut}+\Delta r_{\rm buff})^3$ (or $N^2(r_{\rm cut}+\Delta r_{\rm
  buff})^3$), $\Delta t_{\rm soft} \propto N^{-1/3}$ and $r_{\rm cut}
\propto N^{-1/4}$ and both $N_{\rm int, hard} $ and $N_{\rm int, soft}
$ are proportional to $N^{4/3}$ (or $N^{5/4}$). Here we have neglected
the $\log N$ dependence in the tree part.

This is illustrated in figure \ref{Fig:N-nint}, where we plot $N_{\rm
  int, hard}$ for a plummer sphere as a function of $N$. Following
above discussions, we use the $N$-dependent tree timestep: $\Delta
t_{\rm soft}=(1/256)(N/{\rm 16K})^{-1/3}$ and $N_{\rm int, hard}$ as
well as $N_{\rm int, soft}$ are proportional to $N^{4/3}$.

In figures \ref{Fig:Perfromance} and \ref{Fig:N-de}, we plot the
wall-clock time of execution $T_{\rm cal}$ and the maximum relative
energy errors $|\Delta E_{\rm max}/E_0|$ for the time integration for
10 $N$-body units against $N$. Top (bottom) panel in figure
\ref{Fig:Perfromance} shows the results of the runs with $r_{\rm
  cut}/\Delta t_{\rm soft}$=2 (top panel) and 4 (bottom panel). All
runs in these figures are carried out on NVIDIA GeForce
GTX680\footnote{GTX680 does not have ECC (Error Check and Correct)
  memories. However, as we will see later, we do not observe any large
  energy error in all of our runs, which means the hardware error dose
  not affect our result. Betz, DeBardeleben and Walker
  \cite{2014CCPE...26..2134B} performed Molecular Dynamics
  simulations, in order to investigate the rate of bit-flip error
  events. They observed a single bit-flip error event in about 4700
  GPU*hours without ECC and conclude that the bit-flip error is
  exceedingly rare.}  GPU and Intel Core i7-3770K CPU. For each run,
we use one CPU core and one GPU card.

We also perform the simulations using the direct Hermite integrator
with the same $\eta$ and the standard tree code with the same $\theta$
and $\Delta t_{\rm soft}$. These calculations are performed with the
{\tt Sapporo} GPU library \cite{2009NewA...14..630G} and a standard
tree code with the same $\theta$ and $\Delta t_{\rm soft}$ using the
{\tt Bonsai} GPU library \cite{2012JCoPh.231.2825B}. The calculation
time for our ${\rm P^3T}$ implementation is also proportional to
$N^{4/3}$, as we presented in section \ref{Sect:Method:cost:model},
while for the Hermite integrator it is proportional to $N^{7/3}$. The
${\rm P^3T}$ scheme is faster than the direct Hermite integrator for
$N > {\rm 16K}$ and when $N$=1M (M=$2^{20}$), the ${\rm P^3T}$ scheme
is about 50 times faster than the direct Hermite scheme. The pure tree
code is slightly faster than the ${\rm P^3T}$ scheme, but the
integration errors are worse by several orders of magnitude (see
figure \ref{Fig:Eerr-long} and \ref{Fig:N-de}).

\subsection{Examples of practical applications}\label{Sect:Applicataion}

In sections \ref{Sect:accuracy} and \ref{Sect:Method:cost}, we
presented a detailed discussion on the accuracy and performance of our
${\rm P^3T}$ scheme. However, we performed simple simulations, where
the stellar systems are in the dynamical equilibrium. In this section,
we study the performance of our ${\rm P^3T}$ scheme when applied to
more realistic, or more difficult, simulations by comparing the
results of the Hermite scheme. In section \ref{Sect:Applicataion:CC},
we discuss the case of the simulation of star clusters up to core
collapse. In section \ref{Sect:Applicataion:BHB}, we discuss the case
of a galaxy model with massive central black hole binary.

\subsubsection{Star cluster down to core collapse}\label{Sect:Applicataion:CC}

In this section, we discuss the performance of our ${\rm P^3T}$ scheme
for the simulation of the core collapse of a star cluster. In section
\ref{Sect:Applicataion:CC:IC}, we describe the initial condition and
parameters of the integration scheme. In section
\ref{Sect:Applicataion:CC:results} we compare the calculation results
obtained by the ${\rm P^3T}$ and Hermite schemes, and in section
\ref{Sect:Applicataion:CC:speed} the calculation speed.

\paragraph{Initial conditions}\label{Sect:Applicataion:CC:IC}

We apply the P$^3$T scheme to the evolution of a star cluster
consisting of 16K stars to the moment of the core collapse
\cite{1980MNRAS.191..483L}. We use an equal-mass plummer model as an
initial density profile and we adopt $\eta=0.1$.  We apply the Plummer
softening $\epsilon = 4/N = 1/4096$. The simulations are terminated
when the core number-density exceeds $10^6$, at which point the mean
interparticle distance in the core is comparable to $\epsilon$. Next,
we set $\theta$. We must chose $\theta$ so that the tree force error
is smaller than the force due to the two-body relaxation.  Hernquist
et al. \cite{1993ApJ...402L..85H} pointed out that, for $\theta=0.5$
with monopole and quadrupole, the tree-force error is much smaller
than the force due to the two-body relaxation. Thus we chose
$\theta=0.4$ with quadrupole as a standard model. For comparison, we
also perfrome a run with $\theta=0.8$.

To resolve the motions of the particles in the core, we impose $\Delta
t_{\rm soft}$ to be smaller than 1/128 of the dynamical time of the
core ($\sim \sqrt{3\pi/16\rho_{\rm core}}$, where $\rho_{\rm core}$ is
the core density). To reduce the calculation cost for the Hermite part
we require $r_{\rm cut} \propto \rho_{\rm core}^{-1/3}$ and set the
initial value of $r_{\rm cut} = 1/64$. We also change $\Delta r_{\rm
  buff} = 3\sigma_{\rm core}\Delta t_{\rm soft}$, where $\sigma_{\rm
  core}$ is the velocity dispersion in the core, and $\Delta t_{\rm
  max}=\Delta t_{\rm soft}/4$, as $\Delta t_{\rm soft}$ and
$\sigma_{\rm core}$ are changing. Here, to calculate $\rho_{\rm core}$
and $\sigma_{\rm core}$, we use the formula proposed by Casertano and
Hut \cite{1985ApJ...298...80C}. The same simulation is repeated using
the fourth-order Hermite scheme with the block timesteps with the same
value of $\eta = 0.1$.

\paragraph{Results}
\label{Sect:Applicataion:CC:results}

In figure\,\ref{Fig:CC} we present the evolution of the core densities
$\rho_{\rm core}$ (top panel) and the core radii $r_{\rm core}$
(bottom panel) for ${\rm P^3T}$ and Hermite schemes. For each scheme,
we perform three runs, changing the initial random seed for generating
the initial conditions of the Plummer model. The behaviors of the
cores for all runs are similar. The differences between two schemes
are smaller than run-to-run variations.

Figure \ref{Fig:CCdE} shows the relative energy errors of the runs
with the same initial seed as functions of the core density (top
panel) and the time (bottom panel). The energy errors of the runs with
${\rm P^3T}$ scheme change randomly, whereas those of the Hermite code
grow monotonically. As a result, the ${\rm P^3T}$ scheme with
$\theta=0.4$ conserves energy better than the Hermite scheme in the
long run. The errors for the ${\rm P^3T}$ scheme with $\theta=0.8$ is
slightly worse than that of the Hermite scheme, but the behavior of
the core are similar with other runs. Thus the choice of $\theta=0.4$
is enough to follow the core collapse simulations.

\paragraph{Calculation speed
}\label{Sect:Applicataion:CC:speed}

Figure \ref{Fig:CCtcal} shows the calculation time of the ${\rm P^3T}$
scheme ($\theta=0.4$) and Hermite scheme on GPU. As shown in this
figure, the calculation time of the ${\rm P^3T}$ scheme is dominated
by the tree (soft) part calculation.

Initially the ${\rm P^3T}$ scheme is much faster than the Hermite
scheme, but after the time when $\rho_{\rm core} \sim 10^4$, the ${\rm
  P^3T}$ scheme is slightly slower than the Hermite scheme because in
the ${\rm P^3T}$ scheme, $\Delta t_{\rm soft}$ is proportional to
$\rho_{\rm core}^{-1/2}$. However, even for the ${\rm P^3T}$ scheme,
the CPU time spent after $\rho_{\rm core}$ reaches $10^4$ is small.
As a result, the calculation time to the moment of the core collapse
of the ${\rm P^3T}$ scheme is smaller than that of the Hermite scheme
by a factor of two.

\subsubsection{Orbital evolution of SMBH binary}\label{Sect:Applicataion:BHB}

In this section, we also discuss the performance of the ${\rm P^3T}$
scheme applied to simulations of a galaxy with a supermassive black
hole (SMBH) binary. In section \ref{Sect:Applicataion:BHB:IC}, we
describe the initial conditions and parameters of the integration
scheme. In section \ref{Sect:Applicataion:BHB:results} we compare the
calculation results obtained by the ${\rm P^3T}$ and Hermite schemes,
and in section \ref{Sect:Applicataion:BHB:speed} the calculation
speed.

\paragraph{Initial conditions and methods}\label{Sect:Applicataion:BHB:IC}

We use the Plummer model with $N$=16K, 128K and 256K as the initial
galaxy model. Two SMBH particles with a mass of 1 \% of that of the
galaxy are placed at the positions ($\pm$ 0.5, 0.0, 0.0) with the
velocities (0.0, $\pm$ 0.5, 0.0). We use three values for the cut off
radius with respect to three different kinds of interactions. For the
interaction between field stars (FSs), we set $r_{\rm cut,FS-FS} =
1/256$.  For the interaction between SMBHs, the force is not split and
$F_{\rm soft}=0$. In other words, the force between SMBHs is
integrated with the pure Hermite scheme. We set the cut off radius
between SMBH and FS $r_{\rm cut, BH-FS}=1/32$ which is large enough
that $\Delta t_{\rm soft}$ is smaller than the Kepler time of a
particle in orbit around the SMBH binary at a distance of $r_{\rm
  cut,BH-FS}$. We use the Plummer softening $\epsilon = 10^{-4}$ for
the interactions between FS-FS and FS-SMBH. For the SMBH-SMBH
interaction, we do not use the softening. The accuracy parameter of
timestep criterion for FS $\eta_{\rm FS}$ is $0.1$, and for SMBH
$\eta_{\rm BH}$ is 0.03. We adopt $\Delta r_{\rm buff} = 3\sigma\Delta
t_{\rm soft}$, $\Delta t_{\rm max}=\Delta t_{\rm soft}/4$ and
$\theta=0.4$.

We use $\Delta t_{\rm soft}=1/1024$ at $T=0$, and as the binary
becomes harder, we decrease $\Delta t_{\rm soft}$ to suppress the
aliasing error of the binary. As a standard model, we set $\Delta
t_{\rm soft}$ to be less than half of the Kepler time of the
SMBH binary $t_{\rm kep}$. Only for $N=$128K, we also perform two
other runs, where $\Delta t_{\rm soft}<$ $t_{\rm kep}/4$ and $t_{\rm
kep}$.

We also perform the same simulations by the Hermite scheme with the
same $\eta_{\rm FS}$ and $\eta_{\rm BH}$.

\paragraph{Results}\label{Sect:Applicataion:BHB:results}

Figure \ref{Fig:BHB} shows the evolution of the semi-major axis (top
panel) and eccentricity (middle panel) of the SMBH binary and the
relative energy error (bottom panel) as functions of time for our
standard models ($\Delta t_{\rm soft}<t_{\rm kep}/2$). The behaviors
of the semi-major axis of the SMBH binary for the runs with the same
$N$ agree very well. The hardening rate of the binary depends on $N$
because of the loss-cone refilling through the two-body relaxation
\cite{1980Natur.287..307B, 2004ApJ...602...93M,
  2005ApJ...633..680B}. The evolution of the eccentricity has large
variation, because this evolution is sensitive to small $N$
fluctuation \cite{2007ApJ...671...53M}. In the cases of $N$=16K with
the Hermite scheme, the relative energy error increases dramatically
after $T=150$ because the binding energy and the eccentricity of the
binary are very high.

Figure \ref{Fig:BHB2} is the same as figure \ref{Fig:BHB} but for
several different values of $\Delta t_{\rm soft}$. Thick solid, dashed
and dotted curves indicate the results for $\Delta t_{\rm soft}<t_{\rm
  kep}/4$, $t_{\rm kep}/2$ and $t_{\rm kep}$, respectively. The
orbital parameters show similar behaviors for all runs. The absolute
value of the energy errors of ${\rm P^3T}$ runs ($\sim 10^{-5}$) are
small compared with the binding energy of SMBH binary, which is
roughly $0.05$.

\paragraph{Calculation speed}\label{Sect:Applicataion:BHB:speed}

Figure \ref{Fig:BHBtcal} shows the calculation time for runs for
several different values of $N$ with $\Delta t_{\rm soft}<t_{\rm
  kep}/2$. Initially, the ${\rm P^3T}$ scheme is much faster than the
Hermite scheme. As the SMBH binary becomes harder, the ${\rm P^3T}$
scheme slows down more significantly than the direct Hermite scheme
does. We can see that $T_{\rm cal}$ of the Hermite scheme is roughly
proportional to $a^{-1}$ for $a^{-1}>300$, whereas that of the ${\rm
  P^3T}$ scheme is roughly proportional to $a^{-5/2}$, because $\Delta
t_{\rm soft}$ is proportional to the Kepler time of the binary
($\propto a^{3/2}$). However, the calculation time for all runs with
the ${\rm P^3T}$ scheme is shorter than that with the Hermite scheme
by $a=1/800$. We can also confirm that as we use more $N$, the ratio
of the calculation time of the ${\rm P^3T}$ scheme to the Hermite
scheme become larger. The reason why the ${\rm P^3T}$ scheme becomes
slower for large $a^{-1}$ is simply that we force the timestep of all
particles to be smaller than the orbital period of the SMBH binary.
For the Hermite scheme, we do not put such constraint. Thus, in the
Hermite scheme, particles far away from the SMBH have the timestep
much larger than the orbital period of the SMBH binary. This large
timestep can cause accuracy problem \cite{2008NewA...13..498N}. With
${\rm P^3T}$, it is possible to apply the perturbation approximation
to $F_{\rm soft}$ between the SMBH binary and other particles. Such a
treatment should improve the accuracy and speed of the ${\rm P^3T}$
scheme when SMBH binary becomes very hard.

In figure \ref{Fig:BHBtcal2}, we plot the calculation time of the hard
and soft parts for the standard model with $N$=128k. We can see that
the soft parts dominate the calculation time.

In figure \ref{Fig:BHBtcal3}, we compare the calculation time for the
runs with various $\Delta t_{\rm soft}$ ( $<t_{\rm kep}$, $t_{\rm
  kep}/2$, $t_{\rm kep}/4$). Since the most of the calculation time is
spent after the binary becomes hard, the calculation time strongly
depends on the criterion of the $\Delta t_{\rm soft}$. From figure
\ref{Fig:BHB2}, the evolution of the orbital parameters for all runs
with the ${\rm P^3T}$ scheme are similar for various $\Delta t_{\rm
  soft}$ criterion.  Thus we could chose larger $\Delta t_{\rm soft}
\gtrsim t_{\rm kep}$ after the binary formation.

\section{Conclusions}\label{Sect:Conclusion}

We described the implementation and performance of the ${\rm P^3T}$
scheme for simulating dense stellar systems. In our implementation,
the tree part is accelerated using GPU. The accuracy and performance
of the ${\rm P^3T}$ scheme can be controlled through six parameters:
$\Delta r_{\rm cut}$, $\Delta r_{\rm buff}$, $\Delta t_{\rm soft}$,
$\Delta t_{\rm max}$, $\eta$ and $\theta$. We find that $\Delta r_{\rm
  buff} \gtrsim 2\sigma\Delta t_{\rm soft}$ is good choice to prevent
non-neighbour particles from entering the neighbour sphere. The
integration errors can be controlled through $\Delta t_{\rm
  soft}/\Delta r_{\rm cut}\sigma$. For $\theta = 0.2$, if we set
$\Delta t_{\rm soft}$ to be less than $0.05\Delta r_{\rm cut} /
\sigma$, the integration error is smaller than the tree force
error. For the Hermite part, if we chose $\eta \lesssim 0.2$, the
errors hardly depend on $\Delta t_{\rm max}$.

From the point of view of the balance of the calculation costs between
the tree and Hermite parts, we derive the optimal set of accuracy
parameters, and found that the calculation cost is proportional to
$N^{4/3}$.

The ${\rm P^3T}$ scheme is suitable for simulating large $N$ stellar
clusters with a high density contrast, such as star clusters or
galactic nuclei. We demonstrate the efficiency of the code and show
that it is able to integrate $N$-body systems to the moment of the
core collapse. We also performed the simulations of the galaxy with
the SMBH binary and found that the ${\rm P^3T}$ scheme can be applied
to these simulations.

Finally, we discuss the possibilities of implementation of two
important effects on star cluster evolution to ${\rm P^3T}$. The first
is an effect of a tidal field which dramatically change the collapse
time and the evaporation time of a star cluster. The tidal field
effect can be included in the soft part.

The other is an effect of the stellar-mass binary. A stellar-mass
binary plays an important role in halting the core collapse. In this
paper, we introduce the Plummer softening and neglect these binary
effect. However, we could treat these effects by integrating
stellar-mass binaries in the hard part.

Our ${\rm P^3T}$ code is incorporated in the AMUSE frameworks and free
for use \cite{2013CoPhC.183..456P, 2013A&A...557A..84P}.


\begin{backmatter}

\section*{Competing interests}
  The authors declare that they have no competing interests.

\section*{Author's contributions}
All authors, MI, SPZ and JM conceived of the study.
MI developed the code, performed all simulations and drafted the manuscript.
SPZ and JM helped to draft the manuscript.
All authors read and approved the final manuscript. 

\section*{Acknowledgements}

We are grateful to Jeroen B\'edorf, for preparations of the GPU
cluster and GPU library. We also thanks to Shoichi Oshino, Daniel
Caputo and Keigo Nitadori for stimulating discussion. This work was
supported by NWO (grants VICI [\#639.073.803], AMUSE [\#614.061.608]
and LGM [\# 612.071.503]), NOVA and the LKBF.


\bibliographystyle{bmc-mathphys} 
\bibliography{references}      




\section*{Figures}

\begin{figure}[ph!]
  \includegraphics[width=80mm]{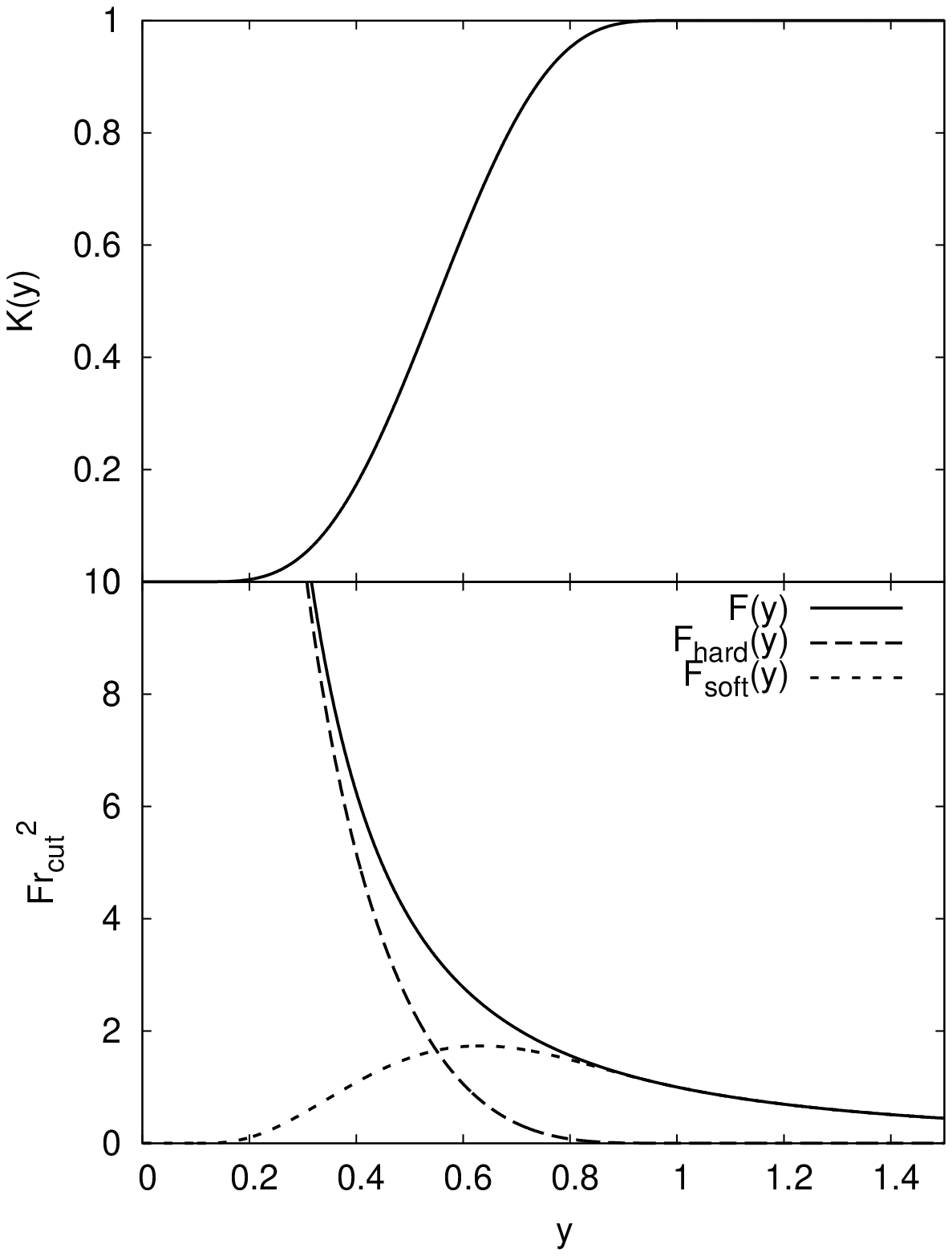}
  \caption{\csentence{The cutoff function $K(y)$ (top) and the
      forces (bottom) as functions of $y=s_{ij}/r_{\rm cut}$.}
    \label{Fig:Ky}
  }
\end{figure}

\begin{figure}[h!]
  \includegraphics[width=100mm]{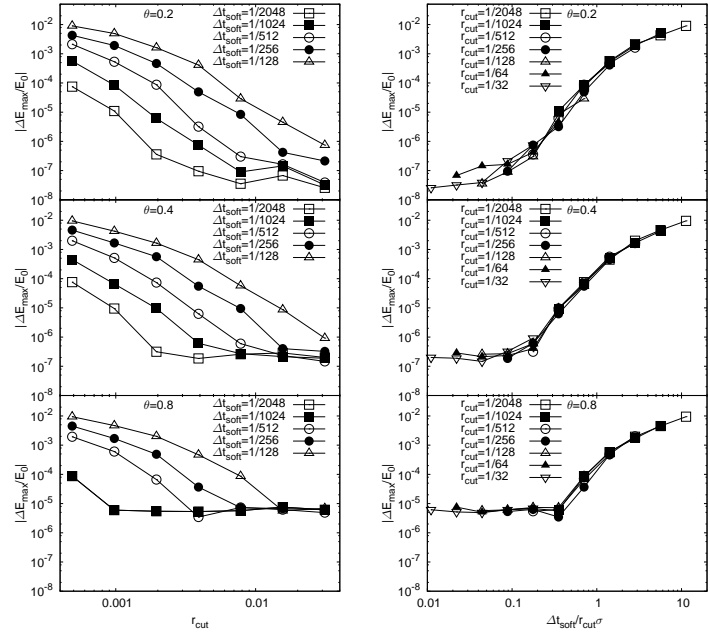}
\caption{\csentence{Maximum relative energy errors as functions of
    $r_{\rm cut}$ (left) and $\Delta t_{\rm soft}/r_{\rm cut}\sigma$
    (right).} Top, middle and bottom panels show the results for
  $\theta$=0.2, 0.4 and 0.8, respectively. For all runs, we use
  $\eta=0.1$, $\Delta t_{\rm max}=\Delta t_{\rm soft}/4$ and $\Delta
  r_{\rm buff}=3 \sigma \Delta t_{\rm soft}$.
\label{Fig:r-de_t-de}
}
\end{figure}

\begin{figure}[h!]
  \includegraphics[width=50mm]{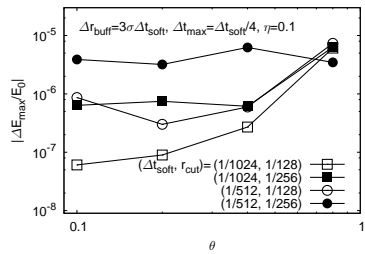}
\caption{\csentence{Maximum relative energy error as a function of
    $\theta$.} For all runs, we use $\eta=0.1$, $\Delta t_{\rm
    max}=\Delta t_{\rm soft}/4$ and $\Delta r_{\rm buff}=3\sigma\Delta
  t_{\rm soft}$.
\label{Fig:th-de}
}
\end{figure}

\begin{figure}[h!]
  \includegraphics[width=50mm]{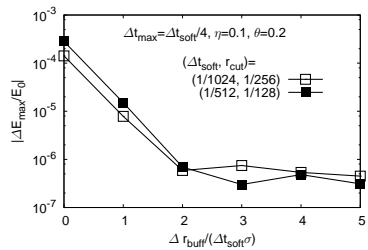}
\caption{\csentence{Maximum relative energy error as a function of
    $\Delta r_{\rm buff}$ in unit of $\Delta t_{\rm soft} \sigma$.}
  Here $\sigma$ is the global three dimensional velocity dispersion of
  the system ($=1/{\sqrt2}$). For all runs, we use $\eta=0.1$, $\Delta
  t_{\rm soft}=1/512$, $t_{\rm max}=\Delta t_{\rm soft}/4$,
  $\theta=0.1$ and $\Delta r_{\rm buff}=3\sigma\Delta t_{\rm soft}$.
\label{Fig:dr-de}
}
\end{figure}

\begin{figure}[h!]
  \includegraphics[width=60mm]{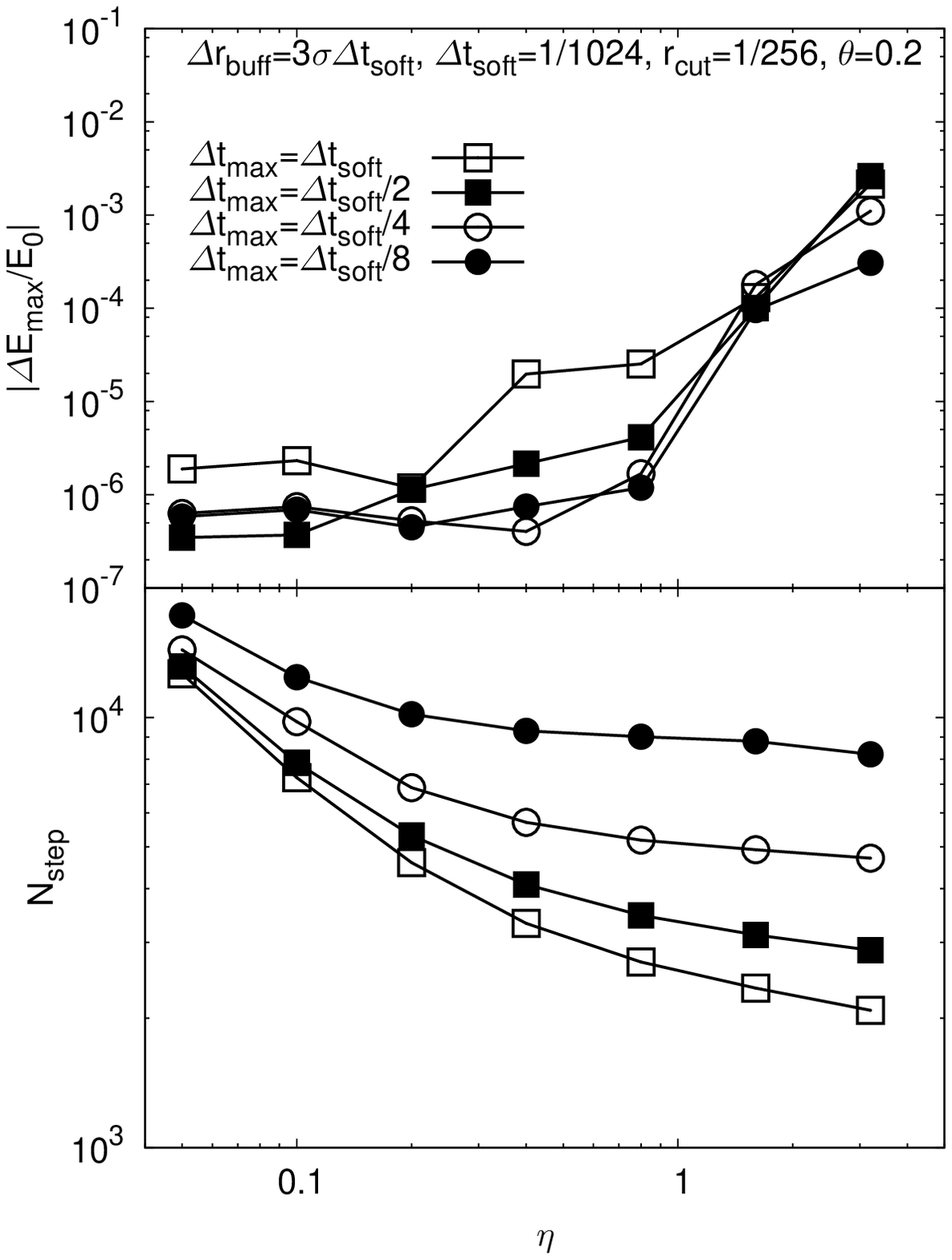}
\caption{\csentence{Maximum relative energy error and the steps for
    the Hermite part against $\eta$.} Top and bottom panels show the
  maximum relative energy error and the steps for the Hermite part par
  particle par unit time against $\eta$, respectively.
\label{Fig:eta-de_eta-nstep}
}
\end{figure}

\begin{figure}[h!]
  \includegraphics[width=100mm]{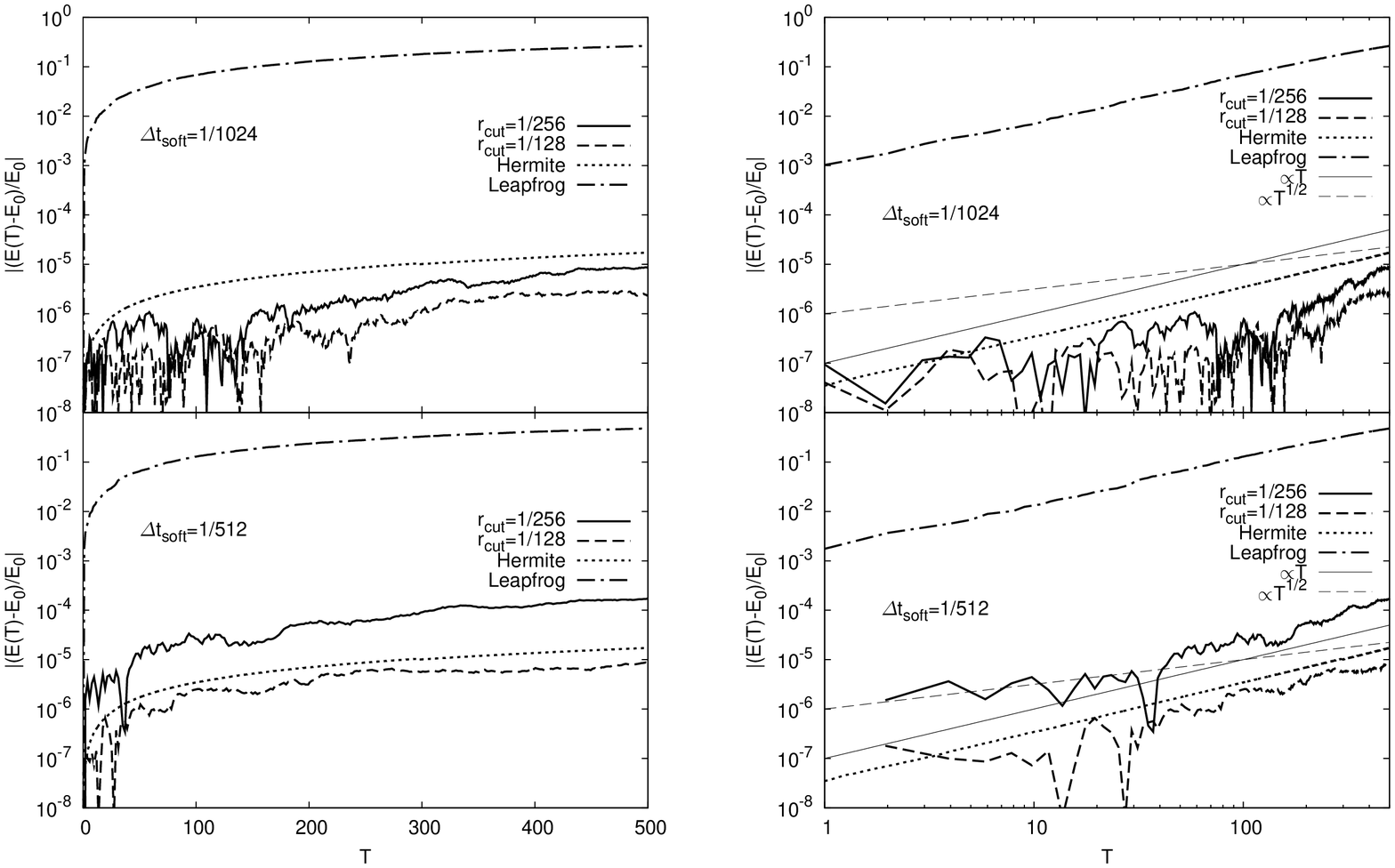}
\caption{\csentence{Evolution of relative energy errors with various
    schemes.} We use $\Delta t_{\rm max}=\Delta t_{\rm soft}/4$,
  $\Delta r_{\rm buff}=3\sigma\Delta t_{\rm soft}$ for all runs and
  $\theta=0.4$ for the tree code and $\eta=0.1$ for the Hermite
  scheme.  In left and right panels, the x-axes are linear and
  logarithmic scales, respectively. Thin curves in right panels are
  proportional to $T$ (solid) and $T^{1/2}$ (dashed).
\label{Fig:Eerr-long}
}
\end{figure}

\begin{figure}[h!]
  \includegraphics[width=60mm]{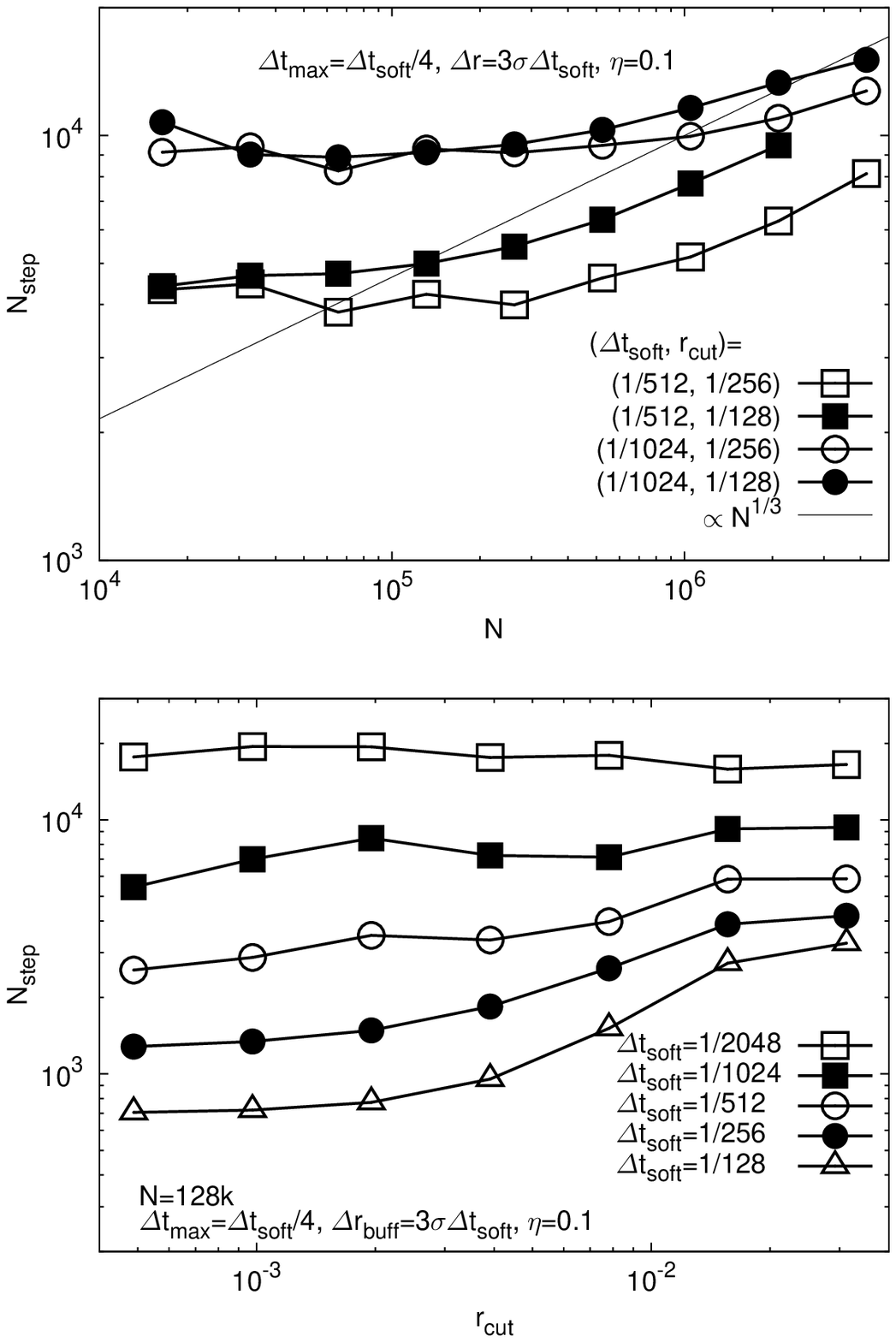}
\caption{\csentence{Number of steps of non-isolated particles as
    functions of $N$ (top) and $r_{\rm cut}$ (bottom).}  Bottom panel
  shows the result of the runs with $N=128$k. For all runs, we chose
  $\eta=0.1$, $\Delta r_{\rm buff}=3\sigma \Delta t_{\rm soft}$ and
  $\Delta t_{\rm max}=\Delta t_{\rm soft}/4$.
\label{Fig:rcut-nstep}
}
\end{figure}

\begin{figure}[h!]
  \includegraphics[width=60mm]{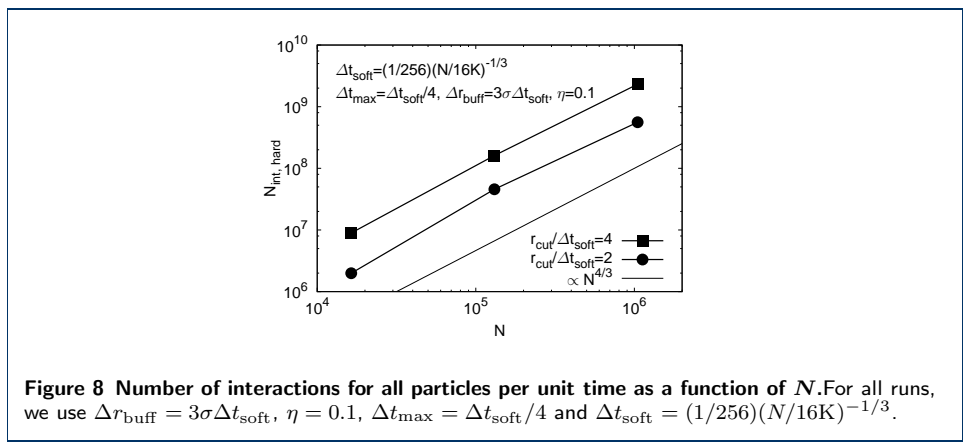}
\caption{\csentence{Number of interactions for all particles per unit
    time as a function of $N$.}For all runs, we use $\Delta r_{\rm
    buff}=3\sigma \Delta t_{\rm soft}$, $\eta=0.1$, $\Delta t_{\rm
    max}=\Delta t_{\rm soft}/4$ and $\Delta t_{\rm soft} = (1/256)
  (N/{\rm 16K})^{-1/3}$.
\label{Fig:N-nint}
}
\end{figure}

\begin{figure}[h!]
  \includegraphics[width=60mm]{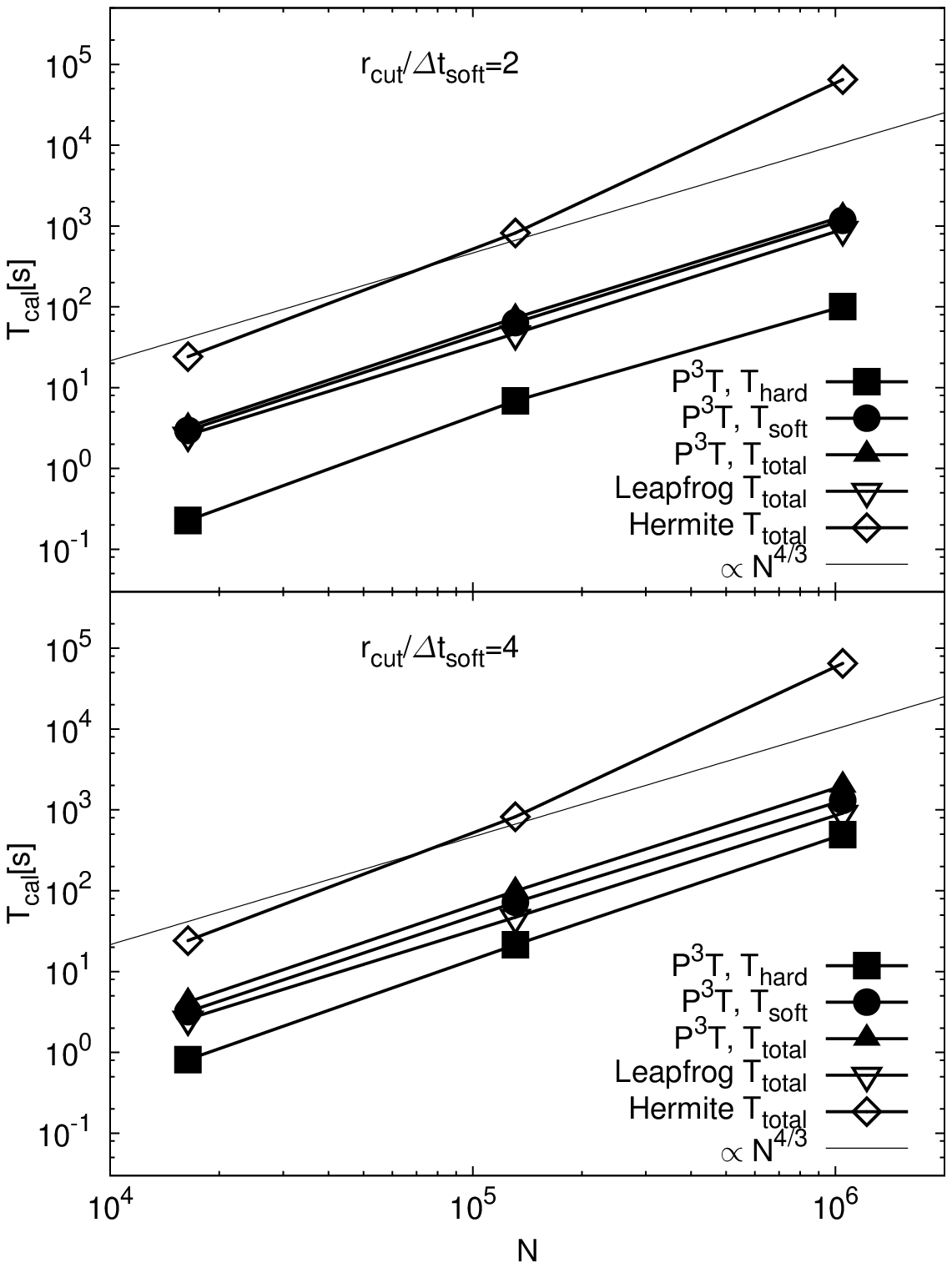}
\caption{\csentence{Wall-clock time of execution as a function of
    $N$.}Top (bottom) panel shows the results of the runs with $r_{\rm
    cut}/\Delta t_{\rm soft} = 2 (4)$. We use $\theta=0.4$,
  $\eta=0.1$, $\Delta r_{\rm buff} = 3\sigma \Delta t_{\rm soft}$ and
  $\Delta t_{\rm soft} = (1/256)(N/{\rm 16K})^{-1/3}$.
\label{Fig:Perfromance}
}
\end{figure}

\begin{figure}[ht!]
  \includegraphics[width=60mm]{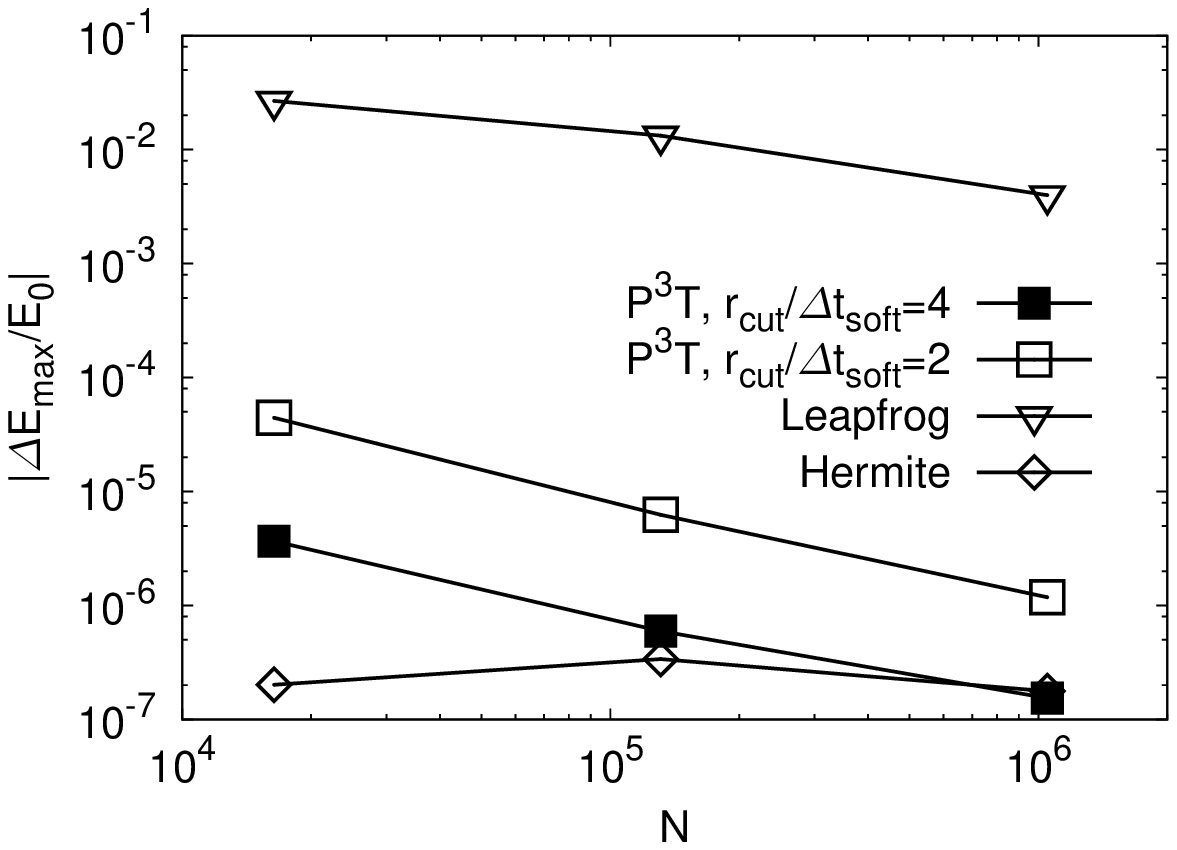}
\caption{\csentence{Maximum relative energy errors over 10 $N$-body
    time units.} All runs are the same as those in figure
  \ref{Fig:Perfromance}.
\label{Fig:N-de}
}
\end{figure}

\begin{figure}[h!]
  \includegraphics[width=50mm]{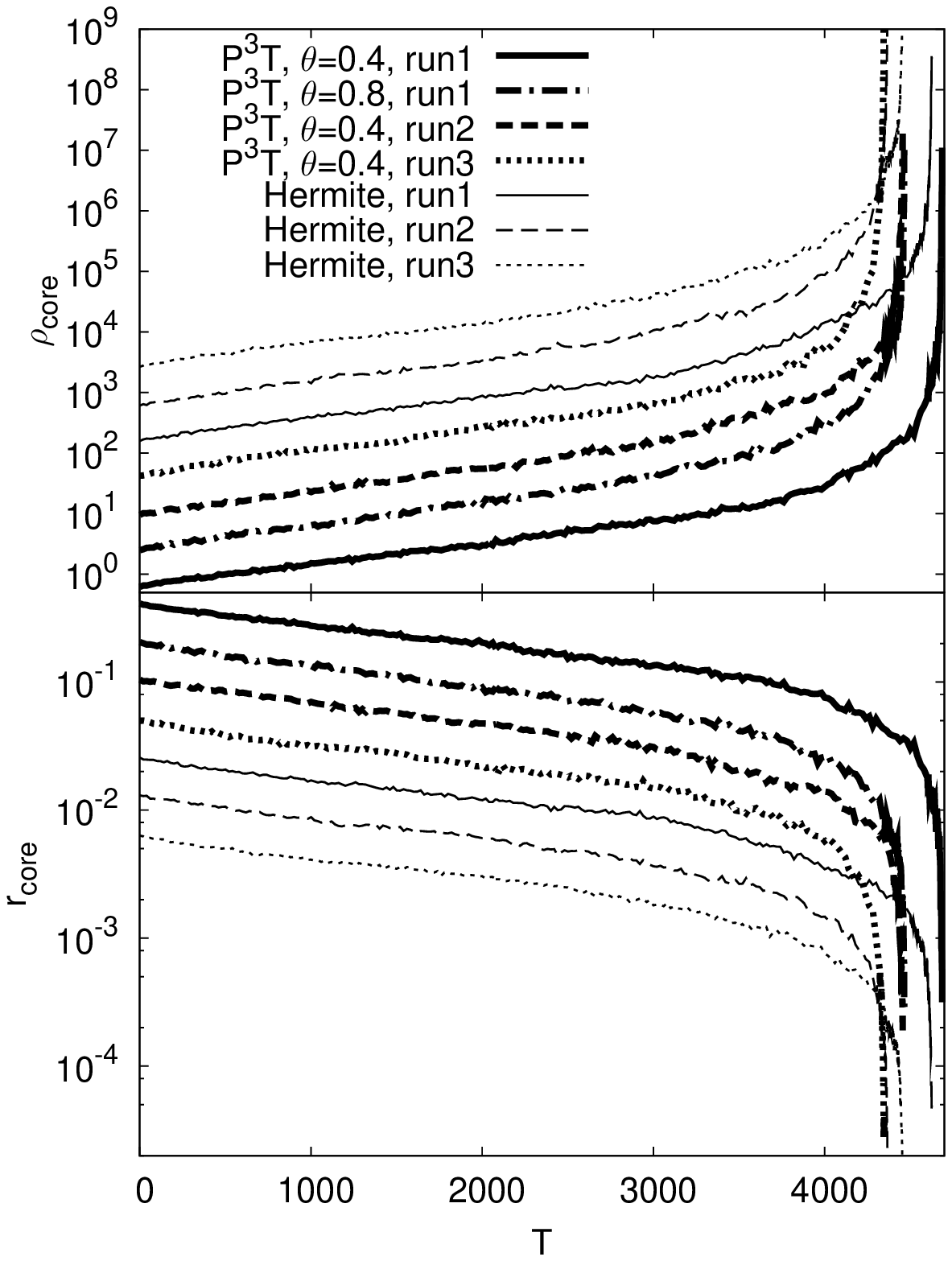}
\caption{\csentence{Time evolution of the core density (top) and the
    core radius (bottom).} Thick and thin curves show the results of
  the ${\rm P^3T}$ and Hermite scheme, respectively. The curves for
  different runs are vertically shifted by a factor of 8 (top) and 2
  (bottom).
\label{Fig:CC}
}
\end{figure}

\begin{figure}[h!]
  \includegraphics[width=60mm]{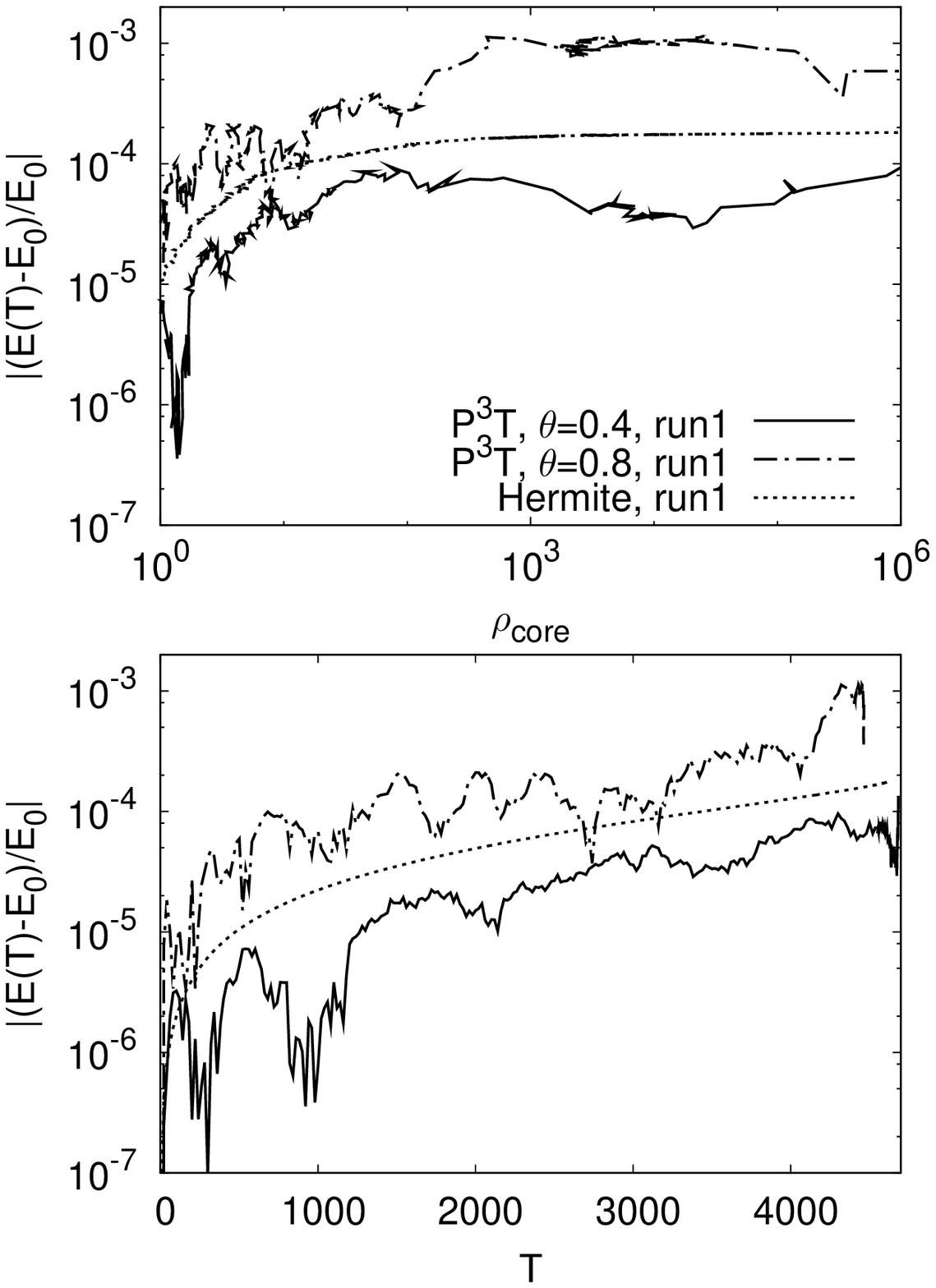}
\caption{\csentence{Relative energy error as functions of $\rho_{\rm
      core}$ (top) and time (bottom).}
\label{Fig:CCdE}
}
\end{figure}

\begin{figure}[h!]
  \includegraphics[width=60mm]{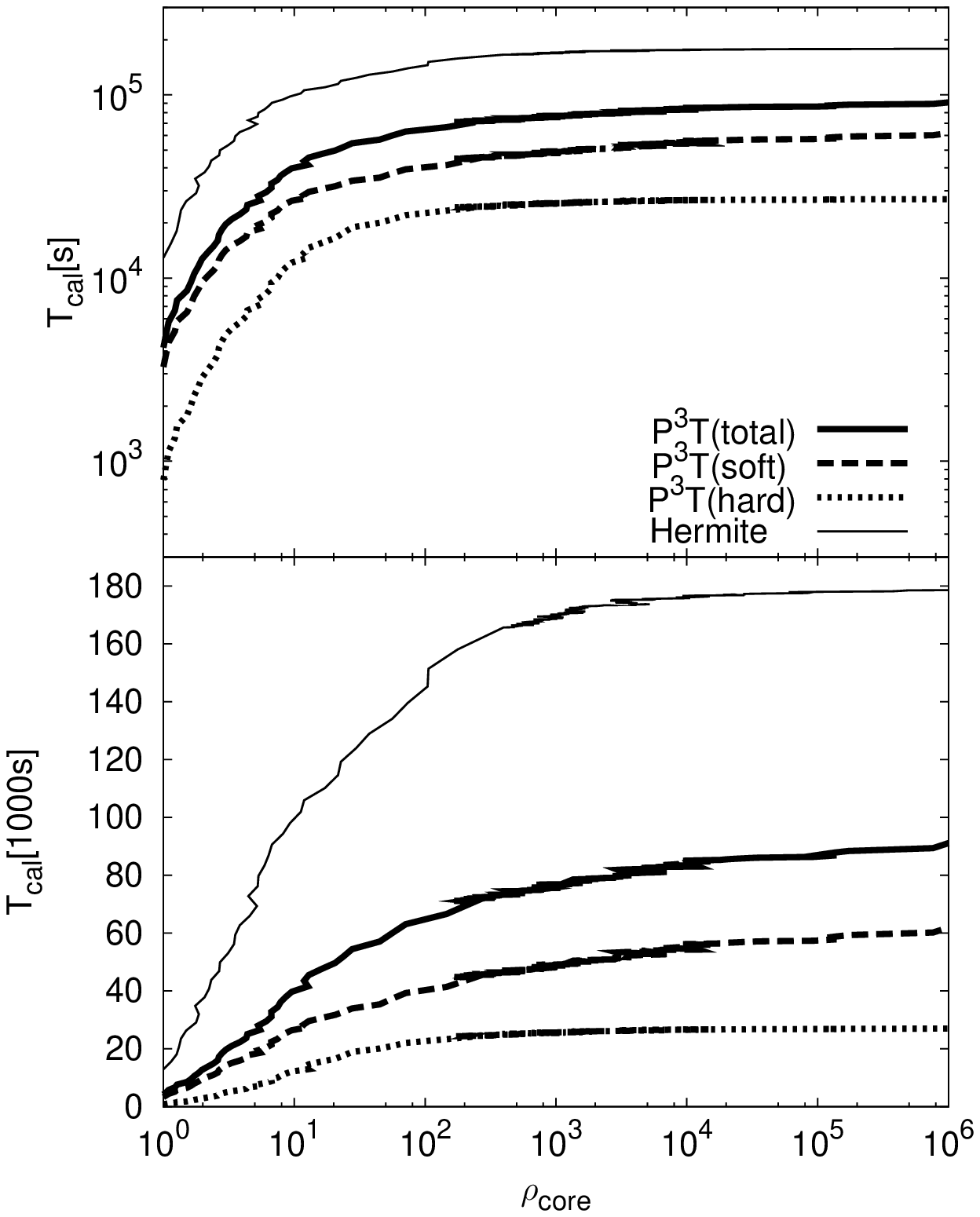}
\caption{\csentence{Wall-clock time of execution as functions of
    $\rho_{\rm core}$.}In top and bottom panels, the y-axes are
  logarithmic and linear scales, respectively.
\label{Fig:CCtcal}
}
\end{figure}

\begin{figure}[h!]
  \includegraphics[width=100mm]{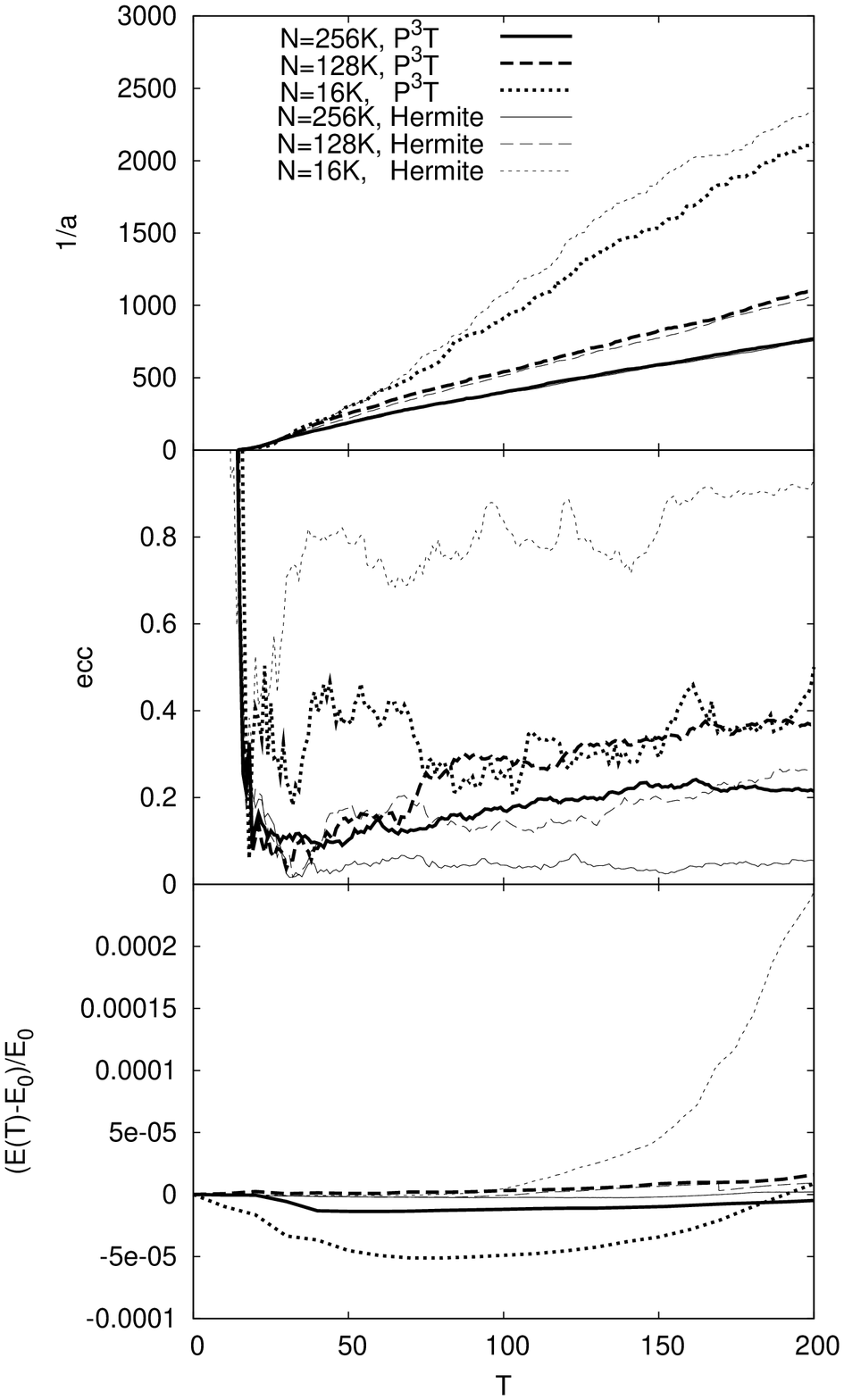}
\caption{\csentence{Evolution of semi-major axis (top), eccentricity
    (middle) of the SMBH binary and energy error (bottom) for several different values of $N$.}
\label{Fig:BHB}
}
\end{figure}

\begin{figure}[h!]
  \includegraphics[width=100mm]{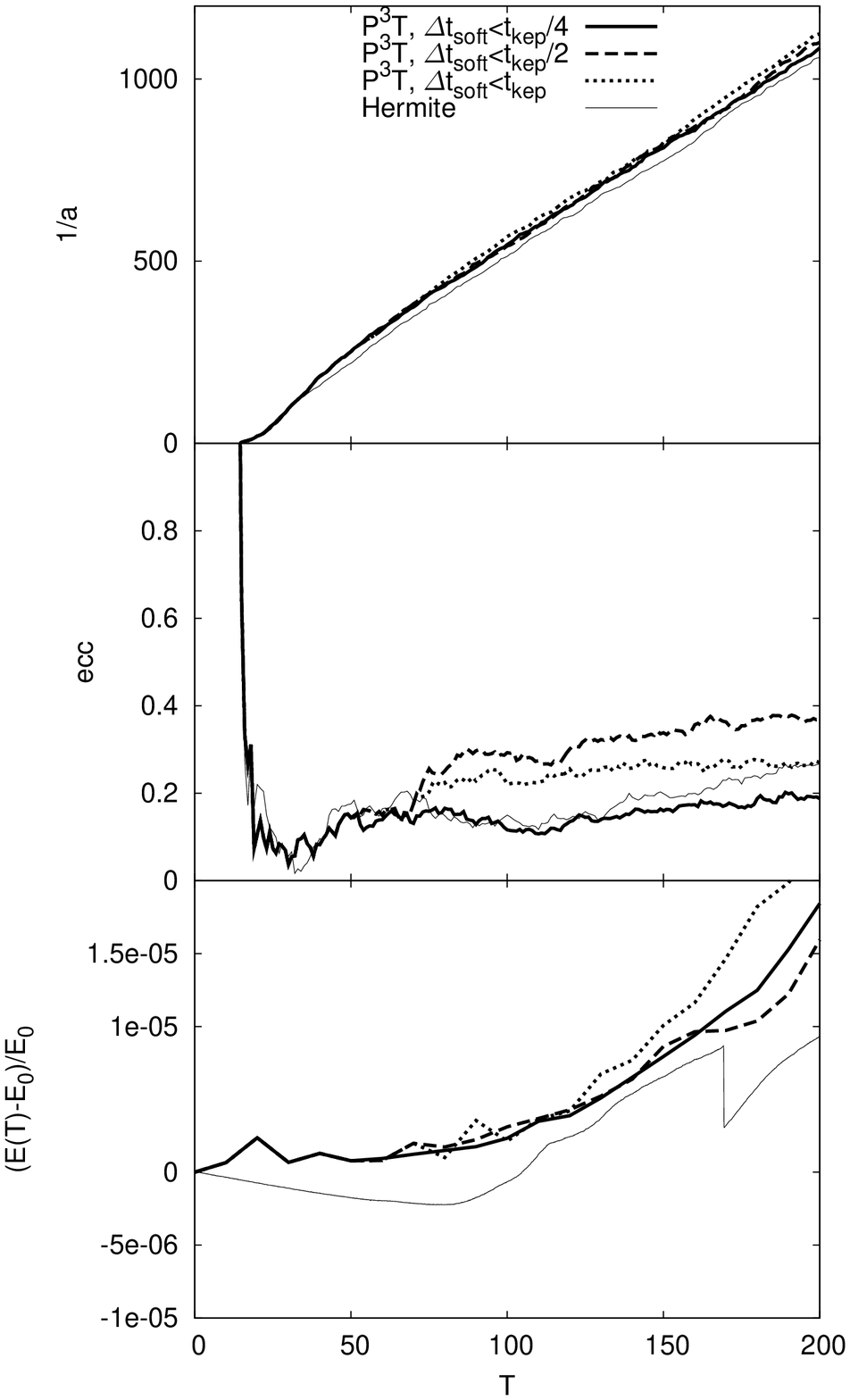}
\caption{
\csentence{Evolution of semi-major axis (top), eccentricity
  	(middle) of the SMBH binary and energy error (bottom) for the several different valuse of $\Delta t_{\rm soft}$.}
\label{Fig:BHB2}
Thick solid, dashed and dotted curves show the results of ${\rm P^3T}$
scheme with $\Delta t_{\rm soft}$ is less than $t_{\rm kep}/4$, $t_{\rm
kep}/2$ and $t_{\rm kep}$, respectively.
}
\end{figure}

\begin{figure}[h!]
  \includegraphics[width=70mm]{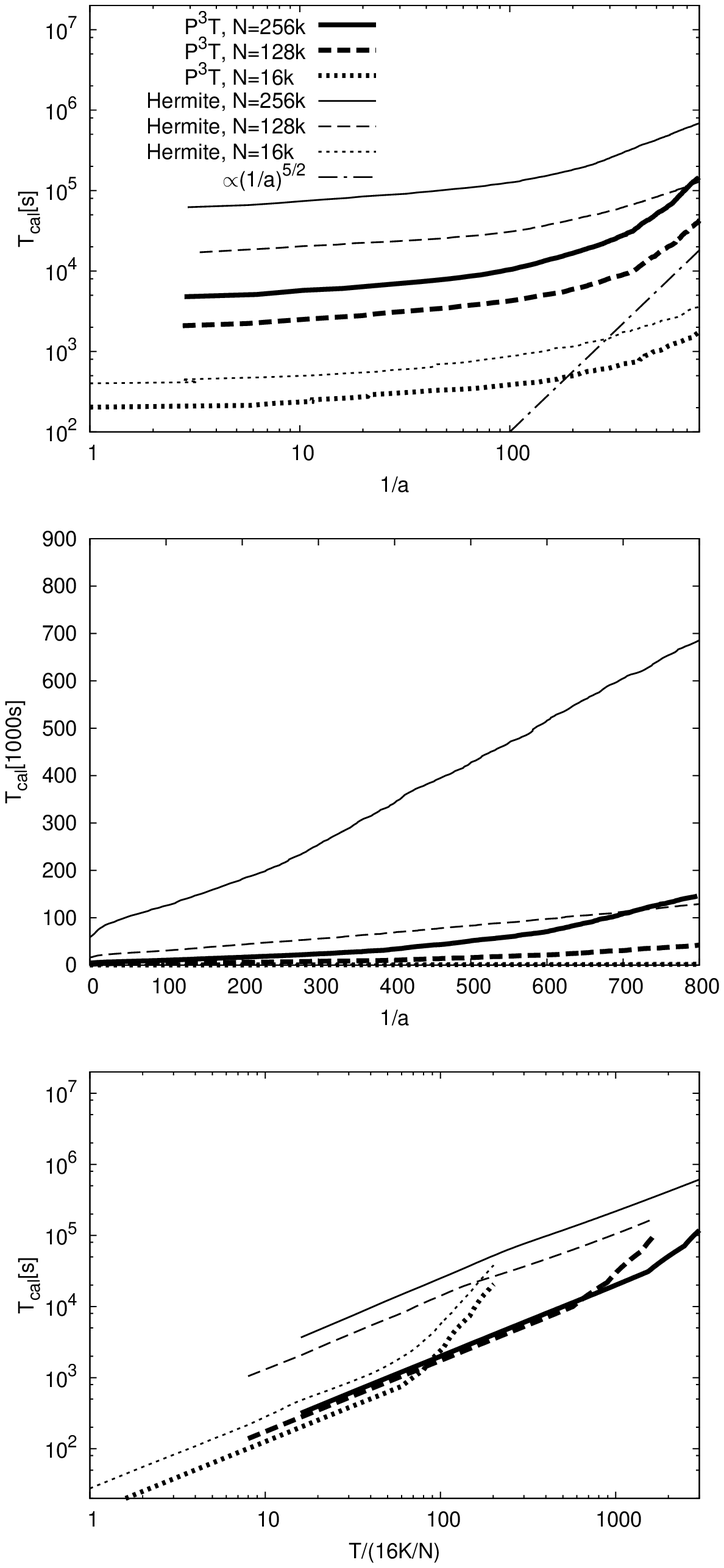}
\caption{
  \csentence{  Wall-clock times as a function of $1/a$ (top and middle) and the
    system time of the simulations (bottom) for several different
    values of $N$.}
  In top and middle panels, the x- and y-axis are
  logarithmic and linear scales, respectively. In bottom panel, x-axis
  is scaled by $N$/16K.
\label{Fig:BHBtcal}
}
\end{figure}

\begin{figure}[h!]
  \includegraphics[width=80mm]{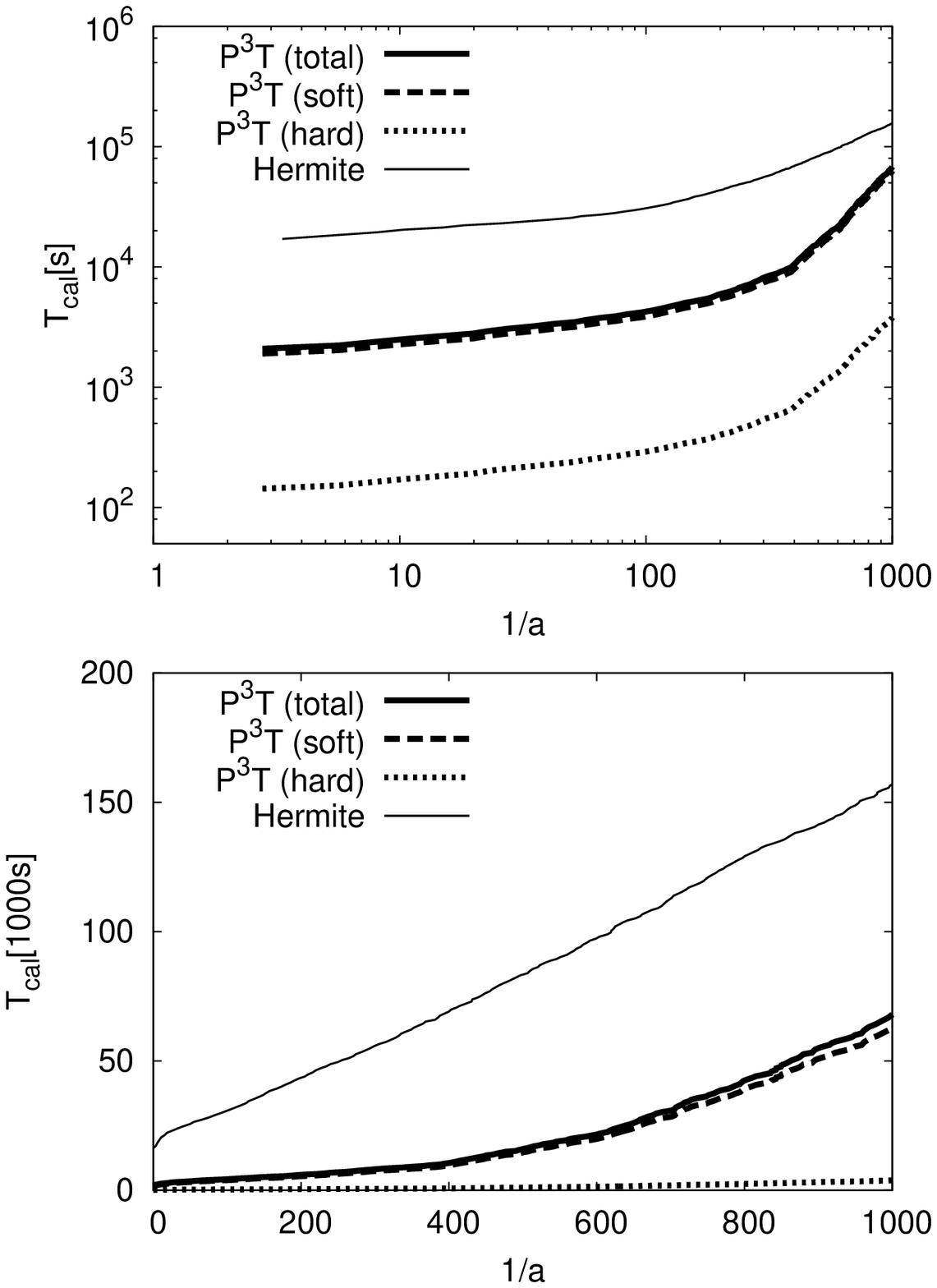}
\caption{\csentence{Wall-clock times as a function of $1/a$.} In top
  and bottom panels, the x- and y-axis are logarithmic and linear
  scales, respectively.
\label{Fig:BHBtcal2}
}
\end{figure}

\begin{figure}[h!]
  \includegraphics[width=80mm]{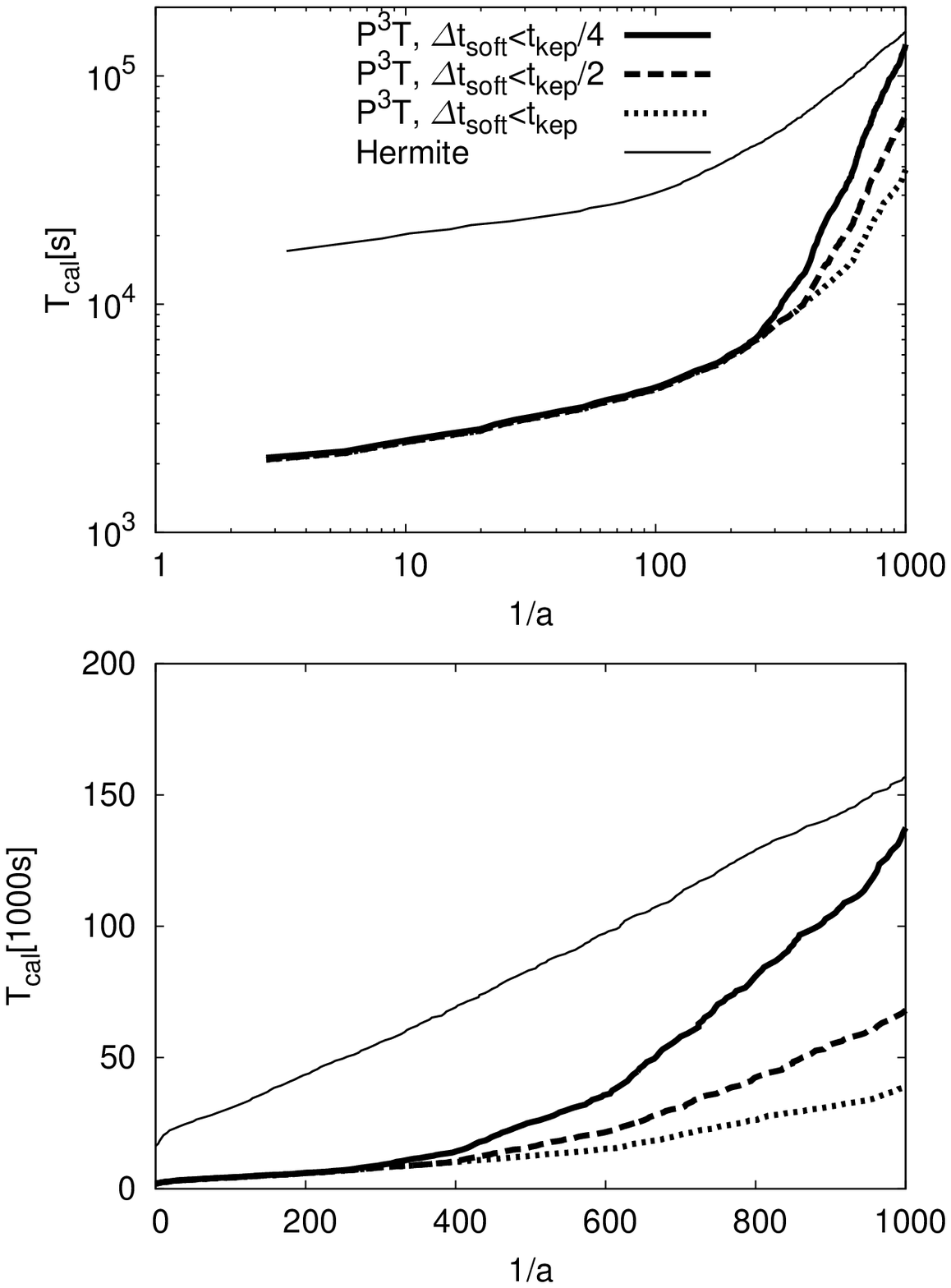} 
\caption{
\csentence{Wall-clock time as a function of $1/a$ for several different values of $\Delta t_{\rm soft}$.} In top
  and bottom panels, the x- and y-axis are logarithmic and linear
  scales, respectively.
\label{Fig:BHBtcal3}
}
\end{figure}

\clearpage

\section*{Tables}
\begin{table}[h!]
\caption{Symbols and definitions for the accuracy parameters of the ${\rm P^3T}$ scheme}
      \begin{tabular}{ll}
        \hline
        $\alpha$        & timestep softening. For all runs, $\alpha = 0.1$. \\
        $\gamma$        & ratio of inner and outer cutoff radius ($r_{\rm in}/r_{\rm cut}$). For all runs, $\gamma = 0.1$.\\
        $\Delta r_{\rm buff}$       & width of the buffer shell. $\Delta r_{\rm buff}=3\sigma\Delta t_{\rm soft}$, as a standard value. \\
        $\Delta t_{\rm soft}$       & timestep of the soft part. $\Delta t_{\rm soft}=(1/256)(N/16{\rm K})^{-1/3}$, as a standard value. \\
        $\Delta t_{\rm max}$        & maximum timestep of the hard part. $\Delta t_{\rm max}=\Delta t_{\rm soft}/4$, as a standard value. \\
        $\epsilon$      & plummer softening length. $\epsilon=(4/N)$, as a standard value. \\
        $\eta$          & accuracy parameter for timestep criterion. $\eta=0.1$, as a standard value.\\
        $r_{\rm cut}$   & outer cutoff radius of smooth transition functions $W$ and $K$. $r_{\rm cut}=4\Delta t_{\rm soft}$, as a standard value. \\
        $r_{\rm in}$    & inner cutoff radius of smooth transition functions $W$ and $K$ ($r_{\rm in}=\gamma r_{\rm cut}$). \\
        $\theta$        & opening criterion for tree. $\theta=0.4$, as a standard value. \\
        \hline
      \end{tabular}
\end{table}

\end{backmatter}
\end{document}